\documentclass[prd,preprint,tightenlines,showpacs,preprintnumbers,nofootinbib,eqsecnum,superscriptaddress]{revtex4-1}

 \usepackage[dvips,final]{graphicx}
  \usepackage{amssymb}
   \usepackage{amsmath}
    \usepackage{amsfonts}
     \usepackage{epsfig}
      \usepackage{bm}

\usepackage[section]{placeins}

\usepackage{multirow}
\usepackage{ctable}
\usepackage{booktabs}
\usepackage{array}
\usepackage{tabularx}
\usepackage{xcolor}
\usepackage{pstricks}

\newcommand{\bl}{\mbox{\boldmath $l$}}
\newcommand{\bp}{\mbox{\boldmath $p$}}
\newcommand{\bq}{\mbox{\boldmath $q$}}
\newcommand{\be}{\mbox{\boldmath $e$}}
\newcommand{\bP}{\mbox{\boldmath $P$}}

\newcommand{\bk}{\mbox{\boldmath $k$}}

\newcommand{\GeV}{\textrm{GeV}}

\newcommand{\ket}[1]{ {#1} \rangle}
\newcommand{\bra}[1]{\langle {#1} }

\begin{document}


\title{The ${\gamma^* \gamma^* \to \eta_c (1S,2S)}$ transition 
form factors for spacelike photons}

\author{Izabela Babiarz}
\email{izabela.babiarz@ifj.edu.pl.pl}
\affiliation{Institute of Nuclear Physics Polish Academy of Sciences, 
ul. Radzikowskiego 152, PL-31-342 Krak{\'o}w, Poland}

\author{Victor P.  Goncalves}
\email{barros@ufpel.edu.br }
\affiliation{Instituto de F\'isica e
Matem\'atica -- Universidade Federal de Pelotas (UFPel) \\
CP 354, CEP 96010-900, Pelotas -- RS -- Brazil}

\author{Roman Pasechnik}
\email{roman.pasechnik@thep.lu.se}
\affiliation{Department of Astronomy and Theoretical Physics,
Lund University, SE-223 62 Lund, Sweden}

\author{Wolfgang Sch\"afer}
\email{Wolfgang.Schafer@ifj.edu.pl} 
\affiliation{Institute of Nuclear
Physics Polish Academy of Sciences, ul. Radzikowskiego 152, PL-31-342 
Krak{\'o}w, Poland}

\author{Antoni Szczurek}
\email{antoni.szczurek@ifj.edu.pl}
\affiliation{Faculty of Mathematics and Natural Sciences,
University of Rzesz\'ow, ul. Pigonia 1, PL-35-310 Rzesz\'ow, Poland\vspace{5mm}}

\begin{abstract}
\vspace{5mm}
We derive the light-front wave function (LFWF)
representation of the $\gamma^* \gamma^* \to \eta_c(1S)\,,\eta_c(2S)$
transition form factor $F(Q_1^2, Q_2^2)$ for two virtual photons 
in the initial state. For the LFWF, we use different models
obtained from the solution of the Schr\"odinger equation for
a variety of $c \bar c$ potentials. We compare our results to 
the BaBar experimental data for the $\eta_c(1S)$
transition form factor, for one real and one virtual photon. 
We observe that the onset of the asymptotic behaviour is strongly delayed
and discuss applicability of the collinear and/or massless limit.
We present some examples of two-dimensional distributions for $F (Q_1^2,Q_2^2)$.
A factorization breaking measure is proposed and factorization breaking 
effects are quantified and shown to be almost model independent.
Factorization is shown to be strongly broken, and a scaling of 
the form factor as a function of $\bar Q^2 = (Q_1^2 + Q_2^2)/2$ is obtained.
\end{abstract}

\maketitle

\section{Introduction}
The description of the hadronic structure in terms of the quark and gluon degrees of freedom is one of the main goals of the Quantum Chromodynamics (QCD). During the last years, our understanding about the partonic distributions has been substantially improved by the experimental data obtained in  $ep$ and $pp$ colliders. Complementary information about the internal structure of mesons can be accessed by the study of the electromagnetic form factors and the meson - photon transition
form factors.  
There has been a lot of interest recently in the exclusive production
of mesons via photon fusion processes studied mainly at the $e^+ e^-$ colliders
\cite{Chernyak:2014wra}. Such studies are strongly motivated by the expectation that at large photon virtualities the measurements of the cross sections will provide strong constrains in the probability amplitude for finding partons in the mesons \cite{Radyushkin:1977gp,Lepage:1979zb,Chernyak:1983ej}.
The meson - photon transition form factors are also of interest because of the role
they play in the hadronic light-by-light contribution to the muon anomalous magnetic moment
\cite{Jegerlehner:2017gek}.

During the last years, a lot of attention has been paid to the case of pseudoscalar light meson - photon transition form factors  \cite{Kroll:2010bf,Stefanis:2014yha}, mainly motivated by the experimental data from the  CLEO, BaBar, Belle and L3 Collaborations  for the $\pi^0$, $\eta$ and $\eta'$ production in $e^+ e^-$ collisions. These collaborations have extracted the transition form factor from single - tag events where only one of the leptons in the final state is measured. In this case, one of the photons is far off the mass shell, while the other is almost real. Such data have allowed to test the collinear factorization approach and the onset of the asymptotic regime, as well motivated the improvement of the theoretical approaches. Similar results have been obtained for the $\eta_c$ production. In this case, the $\eta_c$ mass provides a hard scale that justifies to use a perturbative approach even for zero virtualities. In the past this transition form factor has been studied in different approaches, (although often
only for one virtual photon), such as:
perturbative QCD \cite{Feldmann:1997te,Cao:1997hw}, 
lattice QCD \cite{DE2006,CLQCD}, 
non-relativistic QCD \cite{FJS2015,WWSB2018}, QCD sum rules \cite{LM2012}, as well
as from Dyson-Schwinger and Bethe-Salpeter equations \cite{CDCL2017}. 
In the light-front quark model (LFQM) the case
of one virtual and one real photon has been studied in \cite{GL2013,Ryu:2018egt}.

In the present paper we will treat the heavy meson - photon transition form factor,  focusing our analysis on the pseudoscalar charmonium state
$\eta_c(1S)$ and its radial excitation $\eta_c(2S)$.
Here we will focus on calculating transition form factor 
for both virtual photons, which was not studied so far in the light-front
approach.  These double virtual transition form factors can be measured in $e^+ e^-$ collisions in the double - tag mode, where both electron and positron are detected in the final state. Recent results for the $\eta^{\prime}$ production by the BaBar Collaboration \cite{BaBar:2018zpn} have demonstrated that this study is feasible.
Our study is motivated by the possibility of an accurate measurement of the double virtual transition form factors considering the high luminosity expected at Belle2.
This may open new possibilities, the issue of factorization breaking
of the transition form factors, which will be addressed in this paper. 
Regarding the wave functions of the quarkonia, we wish to use also 
$c \bar c$ wave functions obtained from realistic potential
models. Here we will make use of the solutions obtained in
\cite{CNKP2019}.
We shall investigate how well they can describe the recent BaBar data \cite{Lees:2010de}
for $\gamma \gamma^* \to \eta_c (1S)$.
We shall also calculate transition form factors for 
$\gamma \gamma^* \to \eta_c (2S)$, not yet measured, but could be considered for Belle 2 program. 

The paper is organized as follows. In Section~\ref{Sect:LF-formalism}, we present
the light-front formalism that is used for computations of the $\gamma^* \gamma^* \to \eta_c$ form factor. Here, we provide the details of the light front $\eta_c(1S,2S)$ wave function
calculations in the framework of the Schr\"odinger equation with a chosen set of $c\bar c$ interaction potentials, as well as derive the basic relations for the corresponding 
amplitude and the transition form factor. In Section~\ref{Sect:results}, we discuss the most
relevant numerical results on the $\eta_c$ wave functions and the transition form factors for different $c\bar c$ potentials, which are also compared to the existing BaBar data for the $\eta_c(1S)$ state. Section~\ref{Sect:Conclusions} summarizes the most important results of our analysis.

\section{The light-front formalism for 
the $\gamma^* \gamma^* \to \eta_c$ form factor}
\label{Sect:LF-formalism}

\begin{figure}[t]
\includegraphics[width=8cm]{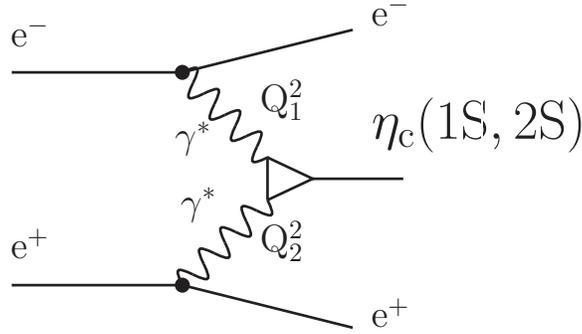}
\caption{The $\eta_c$ production by the interaction of  two virtual photons in $e^+ e^-$ collisions.}
\label{fig:diagram}
\end{figure}

Our goal in this Section is to describe the $\gamma^* \gamma^* \rightarrow \eta_c$ form factor, which can be measured in $e^+ e^-$ collisions using two - photon events in which both photons are far off the mass shell. The typical diagram is represented in Fig. \ref{fig:diagram}. We will consider the light - front formalism, which allows to describe the meson in terms of the quark  degrees of freedom.   
Let us first start from general kinematical considerations.

%
%
The amplitude for the photon fusion $\gamma^* \gamma^* \to \eta_c$ has the general form 
dictated by the $J^{PC} = 0^{-+}$ quantum numbers of the $\eta_c$:
\begin{eqnarray}
{\cal M}_{\mu \nu}(\gamma^*(q_1) \gamma^*(q_2) \to \eta_c) 
= 4 \pi \alpha_{\rm em} \, (-i) \varepsilon_{\mu \nu \alpha \beta} q_1^\alpha q_2^\beta \, F(Q_1^2, Q_2^2) \, .
\end{eqnarray}
Here $Q_i^2 = - q_i^2 \geq 0, i = 1,2$ are the virtualities of photons, 
which we take both to be space-like. 
The form factor $F(Q_1^2, Q_2^2)$ above is the object of interest in this paper.
It is normalized such that the two-photon decay-width of the meson is obtained from
\begin{eqnarray}
\Gamma(\eta_c \to \gamma \gamma) = {\pi \over 4} \alpha^2_{\rm em} M_{\eta_c}^3 \, |F(0,0)|^2 \, .
\label{eq:width}
\end{eqnarray}
For further calculation it is useful to choose a frame in which incoming
photon four-momenta have the form
\begin{eqnarray}
q_1 &=& q_1^+ n^+ + q_{1 \perp}\, , 
q_2 = q_2^- n^- + q_{2 \perp}  \, .
\end{eqnarray}
Here
\begin{eqnarray}
n^{\pm} = {1 \over \sqrt{2}} \, ( 1,0,0,\pm 1) \, ,
\end{eqnarray}
and we will denote the transverse vectors by boldface vectors, e.g.
\begin{eqnarray}
q_{i\perp} = (0,\bq_i,0) \, , \quad q_{i \perp}^2 = - \bq_i^2 \, .
\end{eqnarray}
As the polarization vectors of off-shell photons we will choose
$n^+$ and $n^-$ for the first and second photon,  respectively.
The four-momenta of photons satisfy
$n^+ \cdot q_1 = n^- \cdot q_2 =0$ \footnote{ Notice, that these polarizations are 
precisely the 
polarizations which will dominate in
diagram represented in Fig. \ref{fig:diagram} at high energies. (See e.g. chapter 8 of \cite{Gribov:2009zz}.) 
}
We finally note that the transverse momentum of the meson is $ \bP = \bq_1 + \bq_2$,
and the photon light-front momenta fulfill
\begin{eqnarray}
2 q_1^+ q_2^- = M_{\eta_c}^2 + \bP^2 \, .
\end{eqnarray}
Projected onto the photon polarizations, the amplitude takes the simple form
\begin{eqnarray}
n^{+\mu} n^{-\nu} {\cal M}_{\mu \nu}(\gamma^*(q_1) \gamma^*(q_2) \to \eta_c) = 4 \pi \alpha_{\rm em} \,  (-i) 
[\bq_1, \bq_2] \, F(Q_1^2, Q_2^2) \, ,
\label{eq:FF_def}
\end{eqnarray} 
where $[\bq_1, \bq_2] = q_1^x q_2^y - q_1^y q_2^x$.

\subsection{Light front wave function of $\eta_c$}

We treat the $\eta_c$ as a bound state of a charm quark and antiquark, 
thus assuming that the dominant contribution comes from the $c \bar c$
component in the Fock-state expansion:
\begin{eqnarray}
|\ket{\eta_c; P_+, \bP} = \sum_{i,j,\lambda, \bar \lambda}
{\delta^i_j \over \sqrt{N_c}} \, 
\int {dz d^2\bk \over z(1-z) 16 \pi^3} \Psi_{\lambda \bar \lambda}(z,\bk)
|\ket{c_{i \lambda}(z P_+ ,\bp_c)
\bar c^j_{\bar \lambda}((1-z)P_+,\bp_{\bar c})} + \dots
\nonumber \\
\end{eqnarray}
  
Here the $c$-quark and $\bar c$-antiquark carry a fraction $z$ and $1-z$ 
respectively of the $\eta_c$'s plus-momentum. The light-front helicites 
of quark and antiquark are denoted by $\lambda, \bar \lambda$, and
take values $\pm 1$.
The transverse momenta of quark and antiquark are
\begin{eqnarray}
\bp_c = \bk + z \bP \, , \quad \bp_{\bar c} = -\bk + (1-z) \bP \, .
\end{eqnarray}
The contribution of higher Fock states is expected to be suppressed at large $Q^2$.  Moreover, the fact that non - relativistic potential models are able to describe the quarkonia properties implies that the valence $c\bar{c}$ Fock state probability of $\eta_c$ is almost 1 (see e.g. \cite{Feldmann:1997te}, \cite{Feldmann:1998sh}). In Ref. \cite{GL2013}, the authors have assumed the $\eta_c$ wave function is a combination of the $u\bar{u}$, $d\bar{d}$, $s \bar{s}$ and $c\bar{c}$ Fock states. The resulting predictions for the $\eta_c$ form factor become dependent on the assumptions for the mixing angles, but are similar to those obtained assuming that the $\eta_c$ is a pure $c\bar{c}$ state. Therefore, our assumption for the $\eta_c$ wave function as a pure $c\bar{c}$ state is a very good approximation. The light-front wave function encodes all the necessary 
information on the bound state. Recently there has been a lot 
of interest in calculating light-front wave functions
of heavy quarkonia, see for example \cite{Glazek:2006cu,Gutsche:2014oua,Li:2017mlw}.
Here we follow a different approach which relies on a prescription
due to Terentev \cite{Terentev:1976jk}, valid for weakly bound
non-relativistic systems, which expresses the light-front wave function
in terms of the rest-frame Schr\"odinger wave function.

In the $^{2S+1}L_J = ^1 S_0$ state the wave function of the two-body system 
for canonical spin-projections $\sigma \bar \sigma$ of quark and antiquark has the 
form
\begin{eqnarray}
\Psi_{\sigma \bar \sigma}({\vec p}) = 
{1 \over \sqrt{2}} 
\chi^\dagger_{\sigma} 
i \sigma_2 \chi_{\bar \sigma} 
\, \, \phi(p) \, Y_{00}\Big( {\vec p \over p} \Big) \, =
 {1 \over \sqrt{2}} 
 \chi^\dagger_{\sigma} 
 i \sigma_2 \chi_{\bar \sigma} 
 \, \, {u(p) \over p} \, {1 \over \sqrt{ 4 \pi}}\,
 .
\end{eqnarray}
Here $\chi^\dagger, \chi$ are the Pauli spinors, and $p = |\vec p|$. The normalization condition reads
\begin{eqnarray}
\int d^3 \vec{p}  \sum_{\sigma \bar \sigma} | \Psi_{\sigma \bar \sigma}({\vec p}) |^2 = 1 \, ,
\, \int_0^\infty dp  \, u^2(p) = 1\, .
\end{eqnarray}
In order to apply Terentev's transformation, we first introduce the 
relative momentum $\vec p$ of quark and antiquark in the center - of - mass frame 
at fixed invariant mass $M_{c \bar c}$,
\begin{eqnarray}
\bp = \bk, \quad p_z = (z - {1 \over 2}) M_{c \bar c} \, ,
\label{eq:relative_momentum}
\end{eqnarray}
so that 
\begin{eqnarray}
\vec{p}^2 = {1 \over 4} ( M^2_{c \bar c} - 4 m_c^2) = 
{1 \over 4} \Big( {\bk^2 + m_c^2 \over z (1-z)} - 4 m_c^2 \Big)\, .
\label{eq:running_mass}
\end{eqnarray}
We then decompose the helicity dependent light-front wave function 
(LFWF) into a spin/momentum and radial part as
\begin{eqnarray}
\Psi_{\lambda \bar \lambda} (z,\bk) = 
{\Gamma_{\lambda \bar \lambda}(z,\bk) \over {\cal N}(z,\bk)} \, \phi(z,\bk) \, .
\end{eqnarray}
To obtain the helicity dependent part of the LFWF one needs to transform
the rest-frame  spinors to the light-front spinors, which is
effected by means of a Melosh transform \cite{Melosh:1974cu}. 
This has been done e.g. in Ref.~\cite{Jaus:1989au}, and the
the helicity-dependent vertex reads
\begin{eqnarray}
\Gamma_{\lambda \bar \lambda} (z,\bk) &=& 
\bar u_\lambda(zP_+,\bk) \gamma_5 v_{\bar \lambda}((1-z)P_+,-\bk) \nonumber \\
&=& {1 \over \sqrt{z (1-z)}} \, \Big\{ \lambda m_c \delta_{\lambda - \bar \lambda} 
- \lambda \sqrt{2} \bk \cdot \be(-\lambda) \delta_{\lambda \bar \lambda} \Big\} \, ,
\label{eq:LF-spin-structure}
\end{eqnarray}
where $\be(\lambda) = - (\lambda \be_x + i \be_y)/\sqrt{2}$.
The normalizing function ${\cal N}$ is
\begin{eqnarray}
{\cal N}(z,\bk) = \Big( \sum_{\lambda \bar \lambda} 
\Gamma_{\lambda \bar \lambda}(z,\bk)  
\Gamma^*_{\lambda \bar \lambda}(z,\bk) \Big)^{1/2} 
= \sqrt{2} \, \sqrt{ {\bk^2 + m_c^2 \over z(1-z)}} = \sqrt{2} M_{c \bar c}.
\end{eqnarray}
Then, if we take the meson state to obey the canonical relativistic normalization
\begin{eqnarray}
\bra{\eta_c; P_+', \bP'}|\ket{\eta_c; P_+, \bP} = 2P_+ (2\pi)^3 \delta(P_+'-P_+) \delta^{(2)}(\bP' -\bP), 
\end{eqnarray}
the radial light-front wave function $\phi(z,\bk)$ will be normalized as
\begin{eqnarray}
\int_0^1 { dz \over z(1-z)} \int {d^2 \bk \over 16 \pi^3} \, |\phi(z,\bk)|^2 = N_{c \bar c} = 1\, .
\label{eq:LF-norm}
\end{eqnarray}
To relate the radial LFWF to the rest-frame wave function, we should still take into account a jacobian 
from changing the integration measure
\begin{eqnarray}
{dz d^2 \bk \over z (1-z)} = 4 {d^3\vec{p} \over M_{c \bar c} } \, ,
\end{eqnarray}
so that we obtain the identification
\begin{eqnarray}
\phi(z,\bk) = \pi \sqrt{M_{c\bar c}}\,  {u(p) \over p} \, .
\end{eqnarray}
To lighten up the notation of the amplitude, we also use 
\begin{eqnarray}
\psi(z,\bk) = {\phi (z, \bk) \over {\cal N} (z,\bk) } = 
{\pi \over \sqrt{2 M_{c \bar c}}} { u(p) \over p} \, , 
\label{eq:psi}
\end{eqnarray}
which we also refer to as the radial wave function.

Let us now present the details of computation of the radial wave function by means 
of the Sch\"odinger equation with a set of chosen $c\bar c$ interaction potentials. 
For reviews on these topics, see e.g.~Refs.~\cite{Eichten:2007qx,Voloshin:2007dx}.

\subsection{Schr\"odinger equation and $c\bar c$ interaction potentials}

The charmonium wave function is found in the quark-antiquark rest frame by solving 
the Schr\"odinger equation which for the radial wave function $\psi(r)$ can be 
written as (for more details see Appendix in Ref.~\cite{CNKP2019})
\begin{eqnarray}
\label{SCH-eq}
\frac{\partial^2 u(r)}{\partial r^2} =
(V_{\rm eff}(r)-\epsilon)u(r)\,, \qquad 
u(r) = \sqrt{4\pi}\,r\psi(r) \,, \qquad 
\int\limits_0^{\infty}|u(r)|^2 dr = 1\,, 
\end{eqnarray}
where
\begin{eqnarray}
V_{\rm eff}(r) = m_c V(r) + \frac{l(l+1)}{r^{2}}\,, \qquad 
\epsilon = m_c\,E\,,
\end{eqnarray}
in terms of the interaction $c\bar c$ potential, $V(r)$. Here, we briefly describe 
several models for $V(r)$ chosen for our analysis.

\begin{itemize}
\item {\it Harmonic oscillator:}
\begin{eqnarray}
V(r) = \frac{1}{2}\,m_c\,\omega^{2}\,r^2 \,,
\end{eqnarray}
where for charmonia we adopt
\begin{eqnarray}
\omega=\frac{1}{2}(M_{2S}-M_{1S}) \simeq 0.3 \, {\rm GeV} \,, \qquad m_c=1.4\, {\rm GeV} \,.
\end{eqnarray}
For such a simple choice of the interaction potential 
one finds an analytic solution of the Schr\"odinger equation (\ref{SCH-eq})
\begin{eqnarray}
u(r) = \exp \biggl[ -\frac{1}{4}\,m_c\,\omega\, r^2\biggr] \,,
\end{eqnarray}
yielding a Gaussian shape of the wave function.
\item {\it Cornell potential \cite{Eichten:1978tg,Eichten:1979ms}:}
\begin{eqnarray}
V(r) = -\frac{k}{r} + \frac{r}{a^2}\,, \quad k=0.52\,, \quad a=2.34\,\mathrm{GeV}^{-1} \,,
\end{eqnarray}
and the charm quark mass is fixed to $m_c=1.84$ GeV.
\item {\it Logarithmic potential \cite{Quigg:1977dd}:}
\begin{eqnarray}
V(r) = - 0.6635\,{\rm GeV} + (0.733\,{\rm GeV})\,\log(r \cdot 1\,{\rm GeV})\,,
\end{eqnarray}
with $m_c=1.5$ GeV. 
\item {\it Effective power-law potential \cite{Martin:1980jx,Martin:1980xh}:}
\begin{eqnarray}
V(r) = - 6.41\,{\rm GeV} + (6.08\,{\rm GeV})\,(r\cdot 1\,{\rm GeV})^{0.106} \,,
\end{eqnarray}
assumes $m_c=1.334$ GeV \cite{Barik:1980ai}.
\item {\it Buchm\"uller-Tye (BT) potential \cite{Buchmuller:1980su}:}
\begin{eqnarray}
V(r) = 
\begin{cases}
\frac{k}{r} - \frac{8\pi}{27}\frac{v(\lambda r)}{r} \,, & r \geq 0.01\,{\rm fm} \\
-\frac{16\pi}{25}\frac{1}{r\,\ln w(r)}
\Big(1 + 2\Big(\gamma_E + \frac{53}{75}\Big)
\frac{1}{\ln w(r)} - \frac{462}{625}\,\,
\frac{\ln\ln w(r)}
{\ln w(r)}\Big) \,, & r < 0.01\,{\rm fm} \,,
    \end{cases}
\end{eqnarray}
where $\gamma_E=0.5772$ is the Euler constant, and the function $v(x)$ is known numerically 
from Ref.~\cite{Buchmuller:1980su}, and
\begin{eqnarray}
w(r) = \frac{1}{\lambda^2_{\rm MS}\,r^2}\,,\quad
\lambda_{\rm MS}=0.509\,{\rm GeV}\,,\quad 
k=0.153\,{\rm GeV}^2\,,\quad \lambda=0.406\,{\rm GeV}\,.
\end{eqnarray}
Here, the charm quark mass is taken to be $m_c=1.48\,{\rm GeV}$.
One notices that the BT potential at small $r$ has a Coulomb-like behaviour, 
while at large $r$ -- a string-like behaviour such that its difference from 
the Cornell potential is mainly at small $r$.
\end{itemize}

\subsection{Light-front representation of the amplitude and transition form factor}

Our calculation follows the standard procedure of perturbative QCD for exclusive processes.
We regard the $\gamma^* \gamma^* \to c \bar c$ transition as a hard, perturbatively
calculable process and convolute the amplitude with the bound state wave-function.
We will assume later, that the charm quark mass $m_c$ by itself is large enough to 
justify perturbation theory and apply our results even in the limit of vanishing
photon virtualities. 
Following \cite{Lepage:1980fj} (see Eq.~(A7) therein), the photon fusion amplitude can be expressed as
\begin{eqnarray}
{\cal M}_{\mu \nu}(\gamma^*(q_1) \gamma^*(q_2) \to \eta_c) &=& {{\rm Tr} \,  \openone_{\rm color} \over \sqrt{N_c}} 
\int {dz d^2\bk \over z(1-z) 16 \pi^3} \nonumber \\ 
&\times& \sum_{\lambda \bar \lambda} 
\Psi^*_{\lambda \bar \lambda} (z,\bk) 
{\cal M}_{\mu \nu}(\gamma^*(q_1) \gamma^*(q_2) \to c_{\lambda}(z,\bp_c)
\bar c_{\bar \lambda}(1-z,\bp_{\bar c})) \, . \nonumber \\
\end{eqnarray} 
Here
\begin{eqnarray}
&&{\cal M}_{\mu \nu}(\gamma^*(q_1) \gamma^*(q_2) \to c_{\lambda}(z,\bp_c)
\bar c_{\bar \lambda}(1-z,\bp_{\bar c})) = 4\pi \alpha_{\rm em}\nonumber \\
&&\bar u_\lambda(zP_+, \bp_c) 
\Big( \gamma_\mu {\hat p_c  - \hat q_1- m_c \over  t - m_c^2} \gamma_\nu + 
\gamma_\nu  {\hat p_c - \hat q_2 - m_c \over  u - m_c^2} \gamma_\mu
\Big) v_{\bar \lambda}((1-z)P_+,\bp_{\bar c}),
\end{eqnarray}
with $\hat p \equiv p^\mu \gamma_\mu$, is the standard Feynman amplitude 
		including the $t$-channel and $u$-channel exchange of a quark.
We can express the denominators of the propagators though LF-variables:
\begin{eqnarray}
 u  - m_c^2 &=& -{1 \over 1-z} [ \bl_A^2 + \mu^2 ] \nonumber \\
 t - m_c^2 &=& -{1 \over z} [ \bl_B^2 + \mu^2].
\end{eqnarray}		
Here we introduced the notation
\begin{eqnarray}
\bl_A &=& \bp_{\bar c} - (1-z) \bq_1 = - \bk + (1-z) \bq_2, \nonumber \\
\bl_B &=& \bp_{c} - z \bq_1 = \bk + z \bq_2,
\end{eqnarray}
and
\begin{eqnarray}
\mu^2 = z(1-z) \bq_1^2 + m_c^2 \, .
\end{eqnarray}				
Contracting the Feynman amplitude with $n_\mu^+ n_\nu^-$ allows us to reduce the amplitude
to a form where only simple spinor products of the form $\bar u \hat n^+ v, \bar u \hat n^- u, \bar v \hat n^- v$ need to be performed. Using the spinors from Ref.~\cite{Lepage:1980fj}, we obtain then
\footnote{We have dropped the terms $\propto \Psi^*_{+-} + \Psi^*_{-+}$, which vanish for the pseudoscalar 
state, see Eq.~(\ref{eq:LF-spin-structure}).}

\begin{eqnarray}
&&n^{+\mu}n^{-\nu} {\cal M}_{\mu \nu}(\gamma^*(q_1) \gamma^*(q_2) \to \eta_c) = 4\pi \alpha_{\rm em} {{\rm Tr} \,  \openone_{\rm color} \over \sqrt{N_c}} (-2) 
\int {dz d^2\bk \over \sqrt{z(1-z)} 16 \pi^3}  \nonumber\\
&&\Big\{ \Big[ {1 \over \bl_A^2 + \mu^2 } - {1 \over \bl_B^2 + \mu^2} \Big] \Big[ i [\bk,\bq_1] 
\Big( \Psi^*_{+-}(z,\bk) - \Psi^*_{-+}(z,\bk) \Big) \nonumber \\
&&-\sqrt{2} m \Big( (\be(-)\bq_1) \Psi_{++}^*(z,\bk)
+ (\be(+)\bq_1) \Psi_{--}^*(z,\bk) \Big) \Big] \nonumber \\
&& + \Big[ {1 -z \over \bl_A^2 + \mu^2 } + {z \over \bl_B^2 + \mu^2} \Big] i[\bq_1,\bq_2] 
\Big( \Psi^*_{+-}(z,\bk) - \Psi^*_{-+}(z,\bk) \Big)
  \Big\}\, .
\nonumber \\ 
\end{eqnarray}
Here, subscripts $\pm$ of the wave functions stand for helicities $\pm {1 \over 2}$ of (anti-)quarks.
Inserting the explicit expressions for the helicity dependent wave functions given in Eq.~(\ref{eq:LF-spin-structure}), we observe that there is a large cancellation between the parallel and
antiparallel helicity configurations, and only the last term $\propto [\bq_1, \bq_2]$ survives.
It is then straightforward to read off our result for the $\gamma^* \gamma^* \to \eta_{c}$ form factor by
comparing to Eq.~(\ref{eq:FF_def}):

\begin{eqnarray}
F(Q_1^2, Q_2^2) =  e_c^2 \sqrt{N_c}  \, 4 m_c 
&\cdot& \int {dz d^2 \bk \over z(1-z) 16 \pi^3} \psi(z,\bk) 
\Big\{ 
{1-z \over (\bk - (1-z) \bq_2 )^2  + z (1-z) \bq_1^2 + m_c^2}
\nonumber \\
&+& {z \over (\bk + z \bq_2 )^2 + z (1-z) \bq_1^2 + m_c^2}
\Big\} \, .
\label{eq:FF}
\end{eqnarray}
and then to perform the integration over the azimuthal angle of $\bk$.
Using
\begin{eqnarray}
\int_0^{2 \pi} {d \phi \over 2 \pi} {1 \over A + B \cos \phi} = { 1 \over \sqrt{A^2 - B^2}},
\end{eqnarray}
we can obtain
\begin{eqnarray}
F(Q_1^2, Q_2^2) =  e_c^2 \sqrt{N_c}  \, 4 m_c 
&\cdot& \int {dz k dk  \over z(1-z) 8 \pi^2} \psi(z,\bk) \nonumber \\
&&\Big\{ {1-z \over \sqrt{ (\bk^2 -m_c^2 - z(1-z)\bq_1^2 - (1-z)^2 \bq_2^2)^2 + 4 \bk^2(m_c^2 + z(1-z) \bq_1^2)} } \nonumber \\
&& + {z \over \sqrt{ (\bk^2 -m_c^2 - z(1-z)\bq_1^2 - z^2 \bq_2^2)^2 + 4 \bk^2(m_c^2 + z(1-z) \bq_1^2)} } 
\Big \}. \nonumber \\ 
\label{eq:FF2}
\end{eqnarray}
This form puts into evidence, that the invariant form factor $F(Q_1^2,Q_2^2)$ is a function
of $Q_1^2 = \bq_1^2$ and $Q_2^2 = \bq_2^2$ only. 
Notice that by the Bose-symmetry, the form factor must be
a symmetric function of $Q_1^2, Q_2^2$. This is evidently not obvious
from the representations Eq.~(\ref{eq:FF}) or Eq.~(\ref{eq:FF2}), as the integrand is manifestly
asymmetric in $\bq_1,\bq_2$. However, as will be demonstrated below by the numerical results, our representation has the required symmetry.
In particular, in the limit of one on-shell photon, one must have, that $F(Q^2) \equiv F(Q^2,0)=F(0,Q^2)$, and
the two different integral representations for $F(Q^2)$ which follow from Eq.~(\ref{eq:FF}) coincide
with the ones found in Ref.~\cite{Ryu:2018egt}, where their equivalence was also demonstrated numerically.

A number of limits of Eq.~(\ref{eq:FF}) are interesting. Firstly,
the value $F(0,0)$ for two on-shell photons is related to the two-photon
decay width by Eq.~(\ref{eq:width}).
First note, that
\begin{eqnarray}
F(0,0) = e_c^2 \sqrt{N_c} \, 4m_c \cdot \int{dz d^2\bk \over z(1-z) 16 \pi^3}  {\psi(z,\bk) \over \bk^2 + m_c^2} \, .
\end{eqnarray}
Let us write this result as an integral over the three-momentum $\vec{p}$ and 
the radial wave function $u(p)$.
\begin{eqnarray}
F(0,0) &=& e_c^2 \sqrt{N_c} 4 m_c \int {4 d^3\vec{p} \over {M_{c \bar c}} 16 \pi^3} \, {\psi(z,\bk) \over \bk^2 + m_c^2} \nonumber \\
&=&  e_c^2 \sqrt{2 N_c} {m_c \over \pi} \int_0^\infty {dp \,  p \, u(p) 
\over \sqrt{M_{c \bar c}^3} (p^2 + m_c^2)} \int_{-1}^1 {d \cos \theta \over 1 - \beta^2 \cos^2 \theta} \nonumber \\
&=& e_c^2 \sqrt{2 N_c} \, {2m_c \over \pi} \, \int_0^\infty {dp \, p  \, u(p) \over 
\sqrt{M_{c \bar c}^3} (p^2 + m_c^2)} \, {1 \over 2 \beta} \log\left({1+\beta \over 1 - \beta}\right)\, ,
\end{eqnarray}
Here we used, that in polar coordinates $\bk^2 = p^2 \sin^2 \theta$, hence
$\bk^2 + m_c^2 = (p^2 + m_c^2) ( 1 - \beta^2 \cos^2 \theta)$,
where we introduced
\begin{eqnarray}
\beta = {p \over \sqrt{p^2 + m_c^2} }\quad ,
\end{eqnarray}
the velocity $v/c$ of the quark in the $c \bar c$ cms-frame. Similar results for the
relativistic corrections to the decay width exist in the literature, see e.g. Ref. \cite{Ebert:2003mu}
and references therein. Notice that in these approaches typically the running of
the mass $M_{c \bar c}$ with $p$ is neglected.

In the non-relativistic (NR) limit, where $p^2/m_c^2 \ll 1, \beta \ll 1$, the invariant mass
$M_{c \bar c}$ approaches $M_{c \bar c} = 2 m_c$. If we neglect the binding energy, 
and identify $2 m_c = M_{\eta_c}$, we obtain
\begin{eqnarray}
F(0,0)
&=& e_c^2 \sqrt{N_c} \sqrt{2} {4 \over \pi \sqrt{M_{\eta_c}^5}} \int_0^\infty dp \, p \,  u(p) 
= e_c^2 \sqrt{N_c} {4 \,  R(0) \over \sqrt{\pi M_{\eta_c}^5}} \,\, .
\label{eq:F_00_from_up}
\end{eqnarray}
Here $R(0)$ is the value of the radial wave function $R(r) = u(r)/r$ at the origin.
This yields the well known result for the $\gamma \gamma$-width
(see e.g. Table 2.2 
in Ref. \cite{Novikov:1977dq})
\begin{eqnarray}
\Gamma(\eta_c \to \gamma \gamma) &=& { 4 \alpha_{\rm em}^2 e_c^4 N_c \over M_{\eta_c}^2} |R(0)|^2 \, ,
\label{eq:width_gam_gam}
\end{eqnarray}
which serves as a check on our normalization.
In the same limit, which amounts to an expansion around
$\bk =0$ and $z = 1/2$ in Eq.~(\ref{eq:FF}), we can derive the transition 
form factor in the NRQCD-limit,
\begin{eqnarray}
F(Q_1^2, Q_2^2) = e_c^2 \sqrt{N_c} {4 \over \sqrt{\pi M_{\eta_c}}} {1 \over 
Q_1^2 + Q_2^2 + M_{\eta_c}^2} \, R(0) \, .
\label{eq:NRQCD}
\end{eqnarray}
Still another interesting limit exists, namely at 
very large $Q_i^2$, the $\bk$-smearing becomes unimportant, and one can
neglect $\bk$ in the hard matrix element of Eq.~(\ref{eq:FF}). 
Then only the LFWF appears under the $\bk$ integral, and the hard scattering factorization 
in terms of the distribution amplitude emerges.
We introduce the distribution amplitude (DA) at a scale $\mu_0^2$ as
\begin{eqnarray}
f_{\eta_c} \, \varphi(z,\mu_0^2) &=& {1 \over z(1-z)} {\sqrt{N_c} \, 4 m_c \over 16 \pi^3} 
\int d^2\bk \,  \theta(\mu_0^2 - \bk^2) \,  \psi(z,\bk) \, .
\label{eq:DA}
\end{eqnarray}
The DA is conveniently normalized as
\begin{eqnarray}
\int_0^1 dz \, \varphi(z,\mu_0^2) = 1,
\end{eqnarray}
so that we can extract the so-called decay constant $f_{\eta_c}$ from the
integral over $z$ in Eq.~(\ref{eq:DA}).
The transition form factor simplifies to
\begin{eqnarray}
F(Q_1^2,Q_2^2) = e_c^2 \, f_{\eta_c} \, &\cdot&  \int_0^1 dz \Big\{ {(1-z) \, \varphi(z,\mu_0^2) \over (1-z)^2 Q_1^2 + z(1-z)Q_2^2 + m_c^2} \nonumber \\
&+&{z \, \varphi(z,\mu_0^2) \over z^2 Q_1^2 + z(1-z)Q_2^2 + m_c^2} \, 
\Big\} \, .
\label{eq:collinear_FF}
\end{eqnarray}
This representation is valid in the limit of large photon virtualities $Q_1^2, Q_2^2$.
\section{Numerical results}
\label{Sect:results}
In this Section we will present our results for the doubly virtual transition form factor, which will be calculated for the different models for the meson wave function. A comparison with the current experimental data will be presented and the onset of the asymptotic regime will be discussed. Moreover, in order to estimate the factorization breaking in the transition form factor 
 we will also estimate the normalized form factor, defined by:
\begin{equation}
{\tilde F}(Q_1^2,Q_2^2) = \frac{F(Q_1^2,Q_2^2)}{F(0,0)}
\; ,
\label{normalized_F}
\end{equation}
which nicely quantifies the deviation from point-like coupling.
A popular model for the transition form factor is based
on the vector meson dominance approach (see e.g. Ref.~\cite{Lees:2010de}), and reads
\begin{equation}
    {\tilde F}(Q_1^2,Q_2^2) = \frac{M_{J/\Psi}^{2}}{Q_1^2+M_{J/\Psi}^{2}}
   \cdot \frac{M_{J/\Psi}^{2}}{Q_2^2+M_{J/\Psi}^{2}} \; .
   \label{FF_VDM}
\end{equation}
It features a factorized dependence on the photon virtualities, which
we expect to be broken. In our analysis, we will quantify the  factorization breaking of the transition form factor by estimating the quantity defined by:
\begin{equation}
R(Q_1^2,Q_2^2) = \frac{{\tilde F}(Q_1^2,Q_2^2)}
{{\tilde F}(Q_1^2,0) {\tilde F}(0,Q_2^2)} \; .
\label{factorization_breaking}
\end{equation}
%

\subsection{$c \bar c$ wave functions of $\eta_c(1S)$ and $\eta_c(2S)$}

\begin{figure}
\includegraphics[width = .7\textwidth]{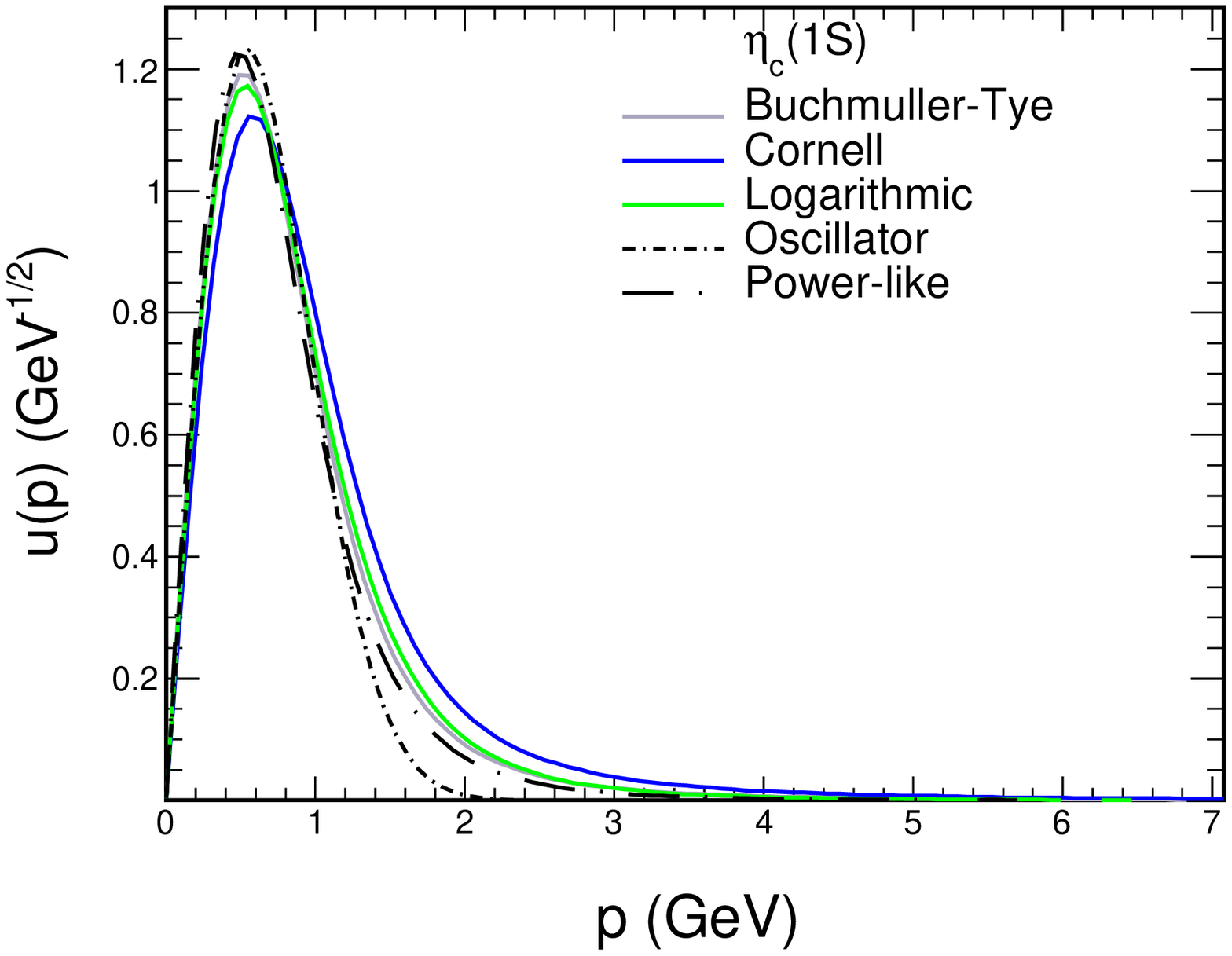}
\includegraphics[width = .7\textwidth]{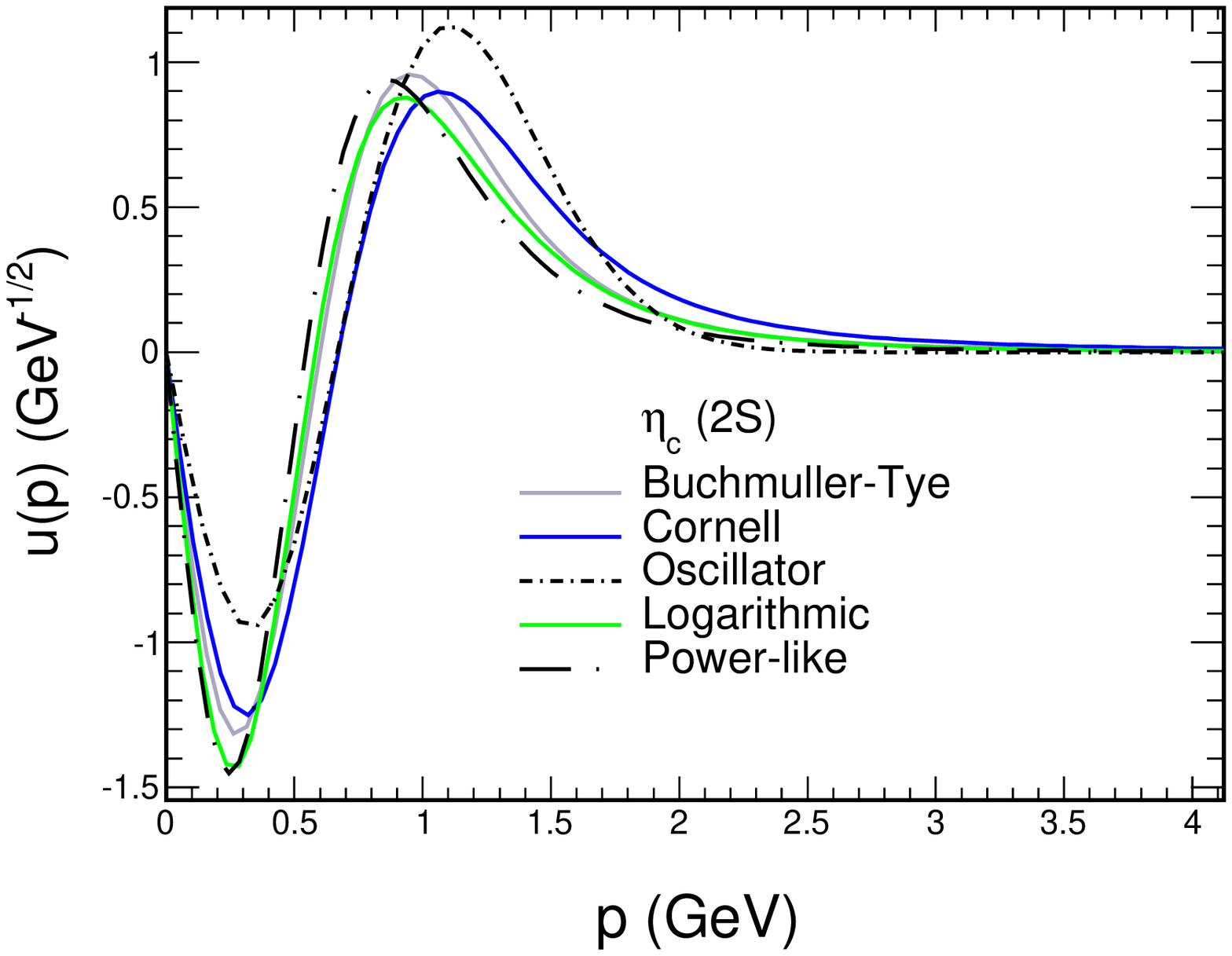}
\caption{Momentum space wave function $u(p)$ for $\eta_c(1S)$ (upper panel) and $\eta_c(2S)$ (lower panel) for different potentials.}
\label{fig:u_p}
\end{figure}

Our wave functions $u(p)$ were obtained by Fourier transform from
the $r$-dependent $c \bar c$ wave functions obtained as a solution of the Schr\"odinger 
equation with different, realistic, potentials from the literature as described 
in the previous Section. In Fig.~\ref{fig:u_p} we show the wave function $u(p)$ for different
potentials for the $1S$ state (left panel) and for the $2S$ radial excitation
(right panel). We have that the different models predict similar shapes for the wave functions, but differ in its predictions for the position of the peaks and the approach to the large momentum limit. 

\begin{figure}
\includegraphics[width=7.5cm]{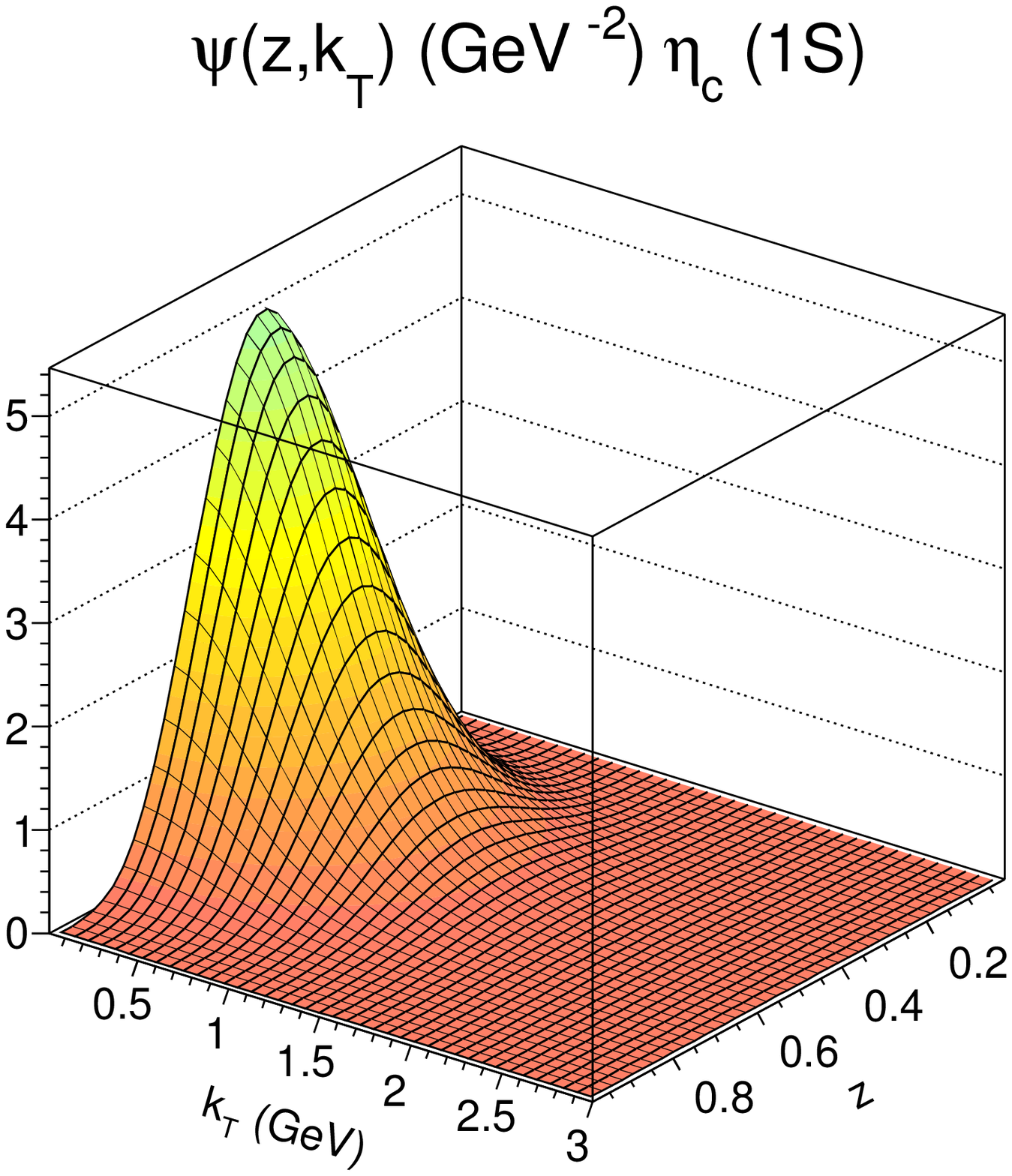}
\includegraphics[width=7.5cm]{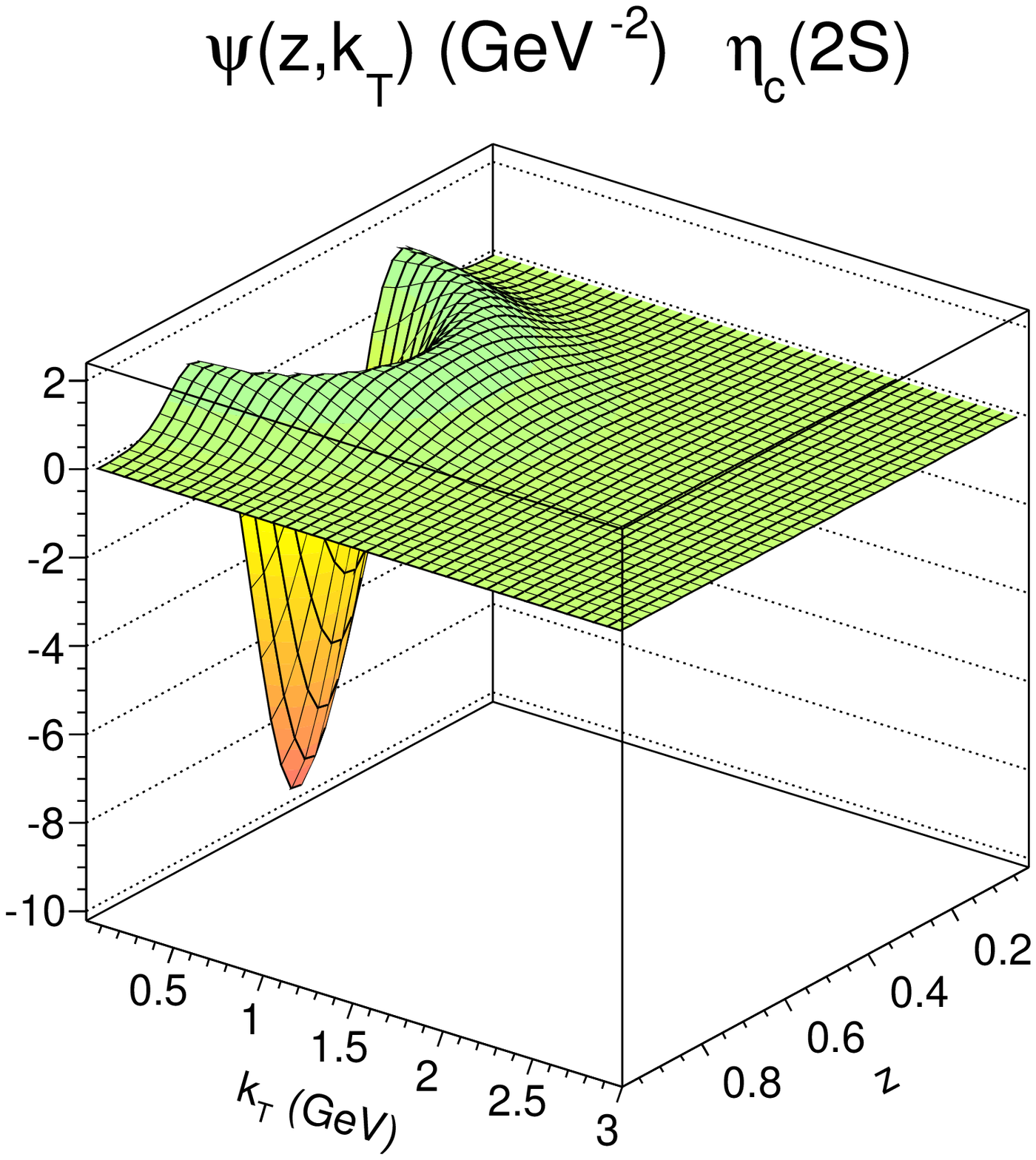}
\caption{The light-front radial wave function $\psi(z,\bk)$ of Eq.~(\ref{eq:psi}) 
in the ($z,k_T$) space for Buchm\"uller-Tye potential for 
$\eta_{c}$(1S) and $\eta_{c}$(2S), respectively. Here $k_T = |\bk|$.}
\label{fig:wf_zkt}
\end{figure}

Using the Terentev prescription (see Eq.~(\ref{eq:relative_momentum})) we obtain 
wave functions in the light-front variables $z$ and $k_T = |\bk|$. 
In Fig.~\ref{fig:wf_zkt} we show as an example the light-front wave function
of Eq.~(\ref{eq:psi}) in the $(z,k_T)$-space for the Buchm\"uller-Tye potential. As expected, the wave function is strongly peaked at $z \approx 1/2$ and is strongly suppressed in the endpoints. 
These properties are shared by all wave
functions for the potentials used by us.
The Cornell potential wave function somewhat stands out as it has the hardest tail at large momenta.

\subsection{$\gamma^* \gamma^*$ transition form factor}

We start the presentation of our results from the value of 
$F(0,0)$ for $\eta_c(1S)$ and $\eta_c(2S)$ obtained from the different
potential model wave functions. 
The obtained results for $F(0,0)$, together with the resulting 
decay width into photons, $\Gamma_{\gamma \gamma}$
are collected in Table~\ref{tab:transition_form_factor_1S}
for the $\eta_c(1S)$
and Table~\ref{tab:transition_form_factor_2S} for the $\eta_c(2S)$,
respectively.
Rather different results are obtained for different potentials, and they
are not always consistent with the one obtained from radiative decay width
$\Gamma_{\gamma \gamma}$, rather well known from recent experiments \cite{Tanabashi:2018oca}. 
It is interesting to compare these --fully relativistic-- results
to the ones obtained from the wave function at the origin
collected in Table \ref{tab:width_from_R_0_1S} and  Table \ref{tab:width_from_R_0_2S} respectively for $\eta_{c}$ (1S) and $\eta_{c}$ (2S). We observe that the relativistic
corrections are fairly strong, especially for the Cornell-potential.

Also shown in Tables  
\ref{tab:transition_form_factor_1S} and \ref{tab:transition_form_factor_2S}
are the values for the so-called decay constant $f_{\eta_c}$.
In the case of the $1S$ state, the agreement with a value 
extracted by the CLEO collaboration \cite{Edwards:2000bb} is generally
quite good. We wish to point out that the decay constant $f_{\eta_c}$ is not
directly related to the two-photon decay width $\Gamma_{\gamma \gamma}$.
Such a relation only exists in the non - relativistic limit, where both quantities 
are expressed in terms of the wave function at the origin.

\begin{figure}
\includegraphics[width=.7\textwidth]{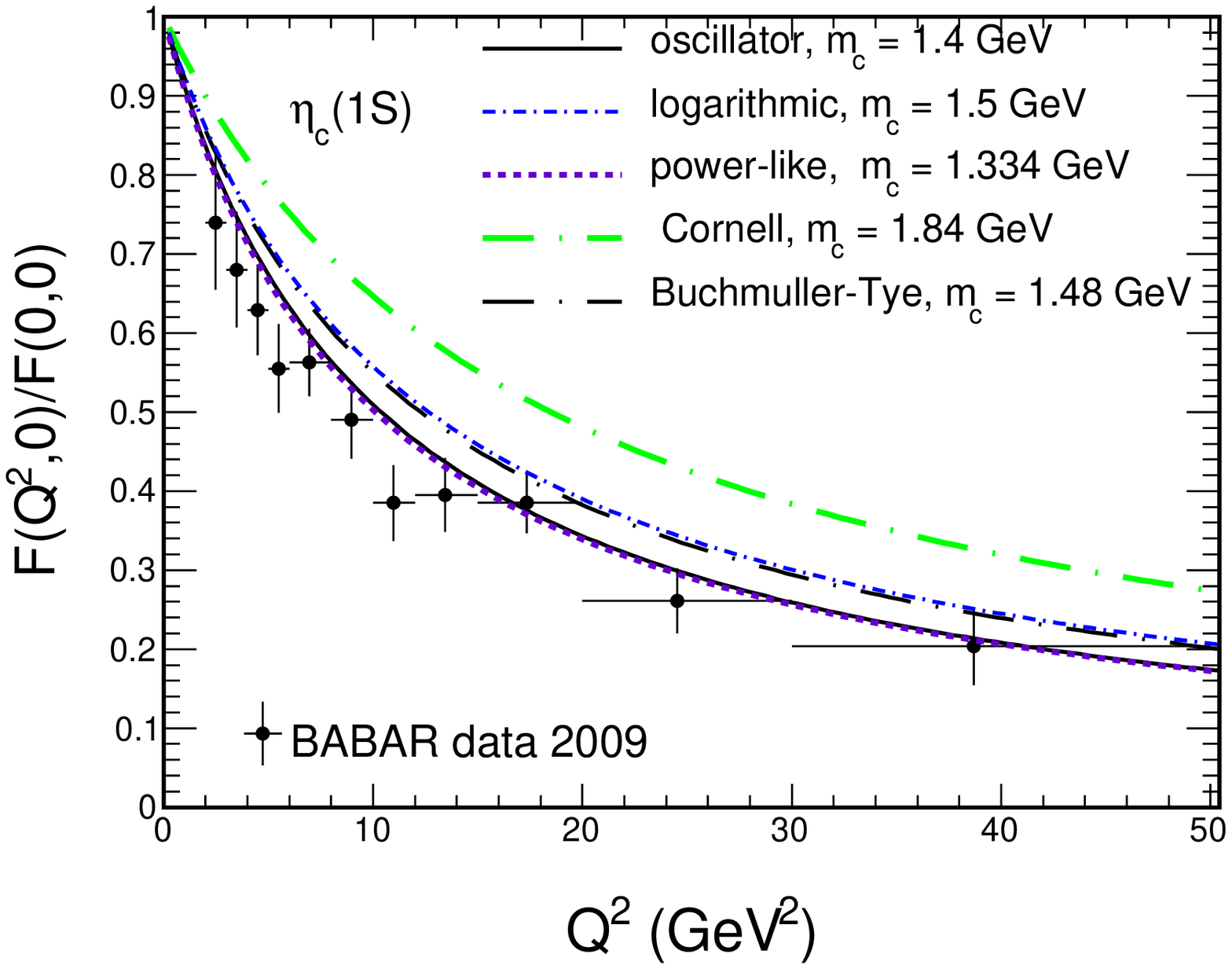}
\includegraphics[width=.7\textwidth]{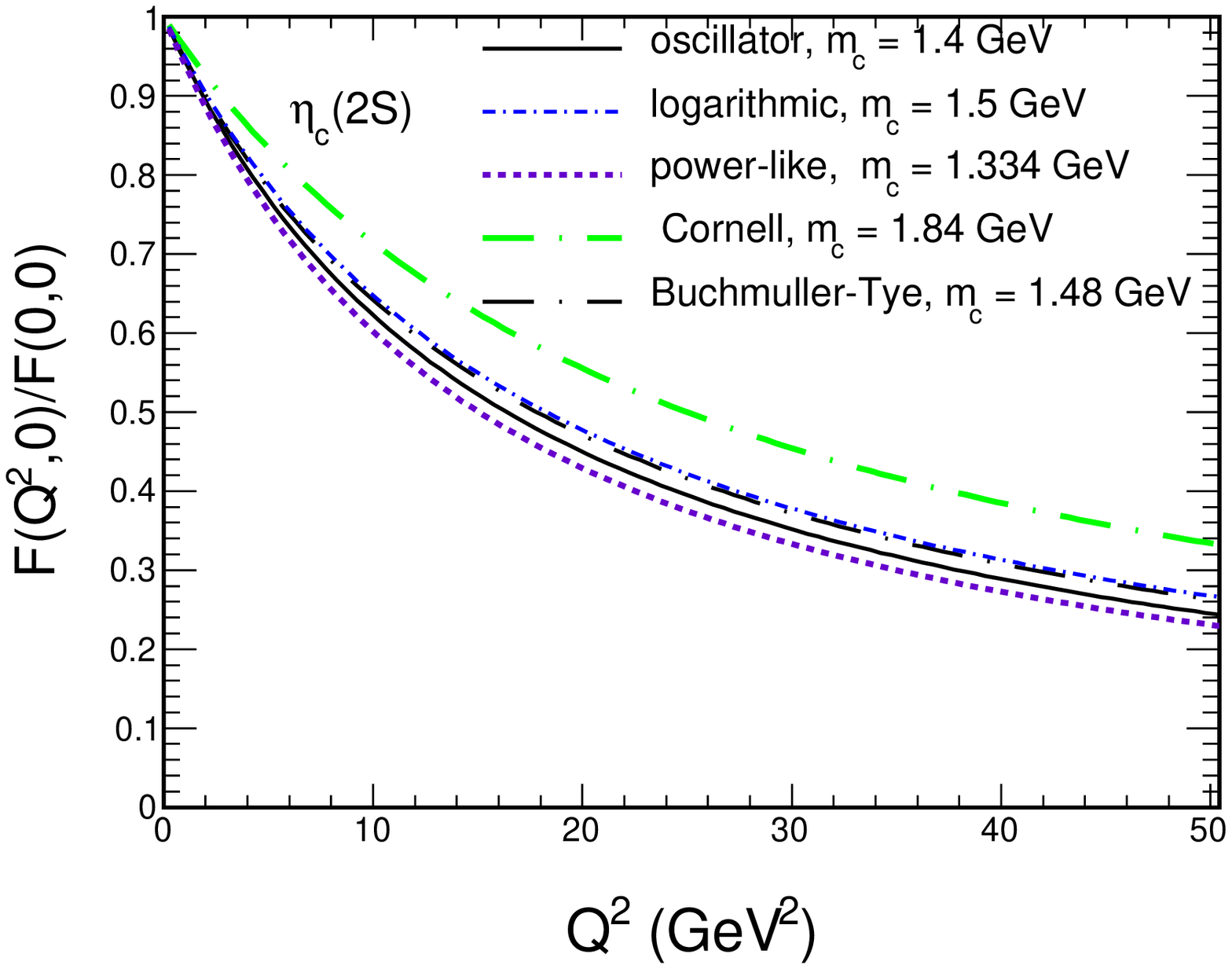}\\
\caption{The dependence of the normalized transition form factor
  $\tilde{F} (Q^2,0)$ on space-like photon virtuality for $\eta_c(1S)$ and
  $\eta_c(2S)$. For the case of the $\eta_c(1S)$, the BaBar experimental
data \cite{Lees:2010de} are shown for comparison.
}
\label{fig:FF_Q2}
\end{figure}

In the upper panel of Fig.~\ref{fig:FF_Q2} we show our results for 
the normalized $\gamma^* \gamma \to \eta_c$ transition form factor ${\tilde F}(Q^2,0)$
as a function of photon virtuality ($Q^2$) for different potential models.
Below we will also use the notation $F(Q^2) = F(Q^2,0) = F(0,Q^2)$, when
		one of the photons is on-shell.
For reference we also show the BaBar experimental data \cite{Lees:2010de}.
The oscillator and power-law potentials give the best description of the BaBar
data. This appears to be related to the lower value of $m_c$ used with
these potentials.
A modification of the quark mass to $m_{c}=1.3\, \GeV$ in the hard
matrix element in fact leads to a much better agreement with the BaBar data.

\begin{table}
\caption{ Transition form factor $|F(0,0)|$ for $\eta_{c}(1S)$ at $Q_1^2=Q_2^2 =$0. }
\label{tab:transition_form_factor_1S}

\begin{tabular}{|l|c|c|c|c|} 
\hline
  potential type &  $m_c \, [\rm{GeV}$]  &  $|F(0,0)| \, [\rm{GeV}^{-1}]
  $ &  $\Gamma_{\gamma \gamma}$\,[$\rm{keV}$]& $f_{\eta_c} [\GeV]$\\
\hline
 harmonic oscillator & 1.4   & 0.051 & 2.89 & 0.2757\\
 logarithmic         & 1.5   & 0.052 & 2.95 & 0.3373\\
 power-like          & 1.334  & 0.059 & 3.87 & 0.3074\\  
 Cornell             & 1.84  & 0.039 & 1.69 & 0.3726\\
 Buchm\"uller-Tye    & 1.48  & 0.052 & 2.95 & 0.3276\\
 \hline
 experiment & - & 0.067 $\pm$ 0.003 \cite{Tanabashi:2018oca} & 5.1
 $\pm$ 0.4 \cite{Tanabashi:2018oca}& 0.335  $\pm$ 0.075 \cite{Edwards:2000bb}\\
\hline
\end{tabular}
\end{table}

\begin{table}
\caption{ Transition form factor $|F(0,0)|$  for $\eta_{c}(2S)$  at $Q_1^2=Q_2^2 =$0. }
\label{tab:transition_form_factor_2S}

\begin{tabular}{|l|c|c|c|c|} 
\hline
  potential type &  $m_c \, [\rm{GeV}$]  &  $|F(0,0)| \, [\rm{GeV}^{-1}]
  $ &  $\Gamma_{\gamma \gamma}$\,[$\rm{keV}$] & $f_{\eta_c} [\GeV]$\\
\hline
 harmonic oscillator & 1.4   & 0.03492 & 2.454 & 0.2530\\
 logarithmic         & 1.5   & 0.02403 & 1.162 & 0.1970\\
 power-like          & 1.334  & 0.02775 & 1.549 & 0.1851\\  
 Cornell             & 1.84  & 0.02159 & 0.938 & 0.2490\\
 Buchm\"uller-Tye    & 1.48  & 0.02687 & 1.453 & 0.2149 \\
 \hline
 experiment \cite{Tanabashi:2018oca} & - & 0.03266 $\pm$ 0.01209 & 2.147 $\pm$ 1.589 & \\
\hline
\end{tabular}

\end{table}

\begin{table}
\caption{ R(0) and $\gamma\gamma$-width for $\eta_c$(1S) derived in the non-relativistic limit.}
\label{tab:width_from_R_0_1S}

\begin{tabular}{|l|c|c|c|} 
\hline
  potential type &  
  R(0)$[\GeV^{3/2}]$ & 
  $\Gamma_{\gamma \gamma}$\,[$\rm{keV}$] $\rm{M}=\rm{M}_{\eta_c}$ & $\Gamma_{\gamma \gamma}$\,[$\rm{keV}$] $\rm{M}=2m_{c}$\\
\hline
 harmonic oscillator & 0.6044  & 5.1848 & 5.8815 \\
 logarithmic         & 0.8919  & 11.290 & 11.157 \\
 power-like          & 0.7620  & 8.2412 & 10.297 \\  
 Cornell             & 1.2065  & 20.660 & 13.568 \\
 Buchm\"uller-Tye    & 0.8899  & 11.240 & 11.409 \\
 \hline
\end{tabular}

\end{table}

\begin{table}
\caption{ R(0) and $\gamma\gamma$-width for $\eta_c$(2S) derived in the non-relativistic limit.}
\label{tab:width_from_R_0_2S}

\begin{tabular}{|l|c|c|c|c|} 
\hline
  potential type & R(0) $[\GeV^{3/2}]$ & 
  $\Gamma_{\gamma \gamma}$\,[$\rm{keV}$] $\rm{M}=\rm{M}_{\eta_c}$ & $\Gamma_{\gamma \gamma}$\,[$\rm{keV}$] $\rm{M}=2m_{c}$\\
\hline
 harmonic oscillator & 0.7402 & 5.2284 & 8.8214\\
 logarithmic         & 0.6372 & 3.8745 & 5.6946\\
 power-like          & 0.5699 & 3.0993 & 5.7594\\  
 Cornell             & 0.9633 & 8.8550 & 8.6493\\
 Buchm\"uller-Tye    & 0.7185 & 4.9263 & 7.4374\\
 \hline
\end{tabular}

\end{table}

The $\gamma^* \gamma$ transition form factor for the $\eta_c(2S)$ is 
shown in the lower panel of Fig. \ref{fig:FF_Q2} .
In this case the ${\tilde F} (Q^2,0) = {\tilde F} (0,Q^2)$ has a 
somewhat harder tail as a function of $Q^2$ than for the $\eta_c (1S)$. 

\begin{figure}
\includegraphics[width=.7\textwidth]{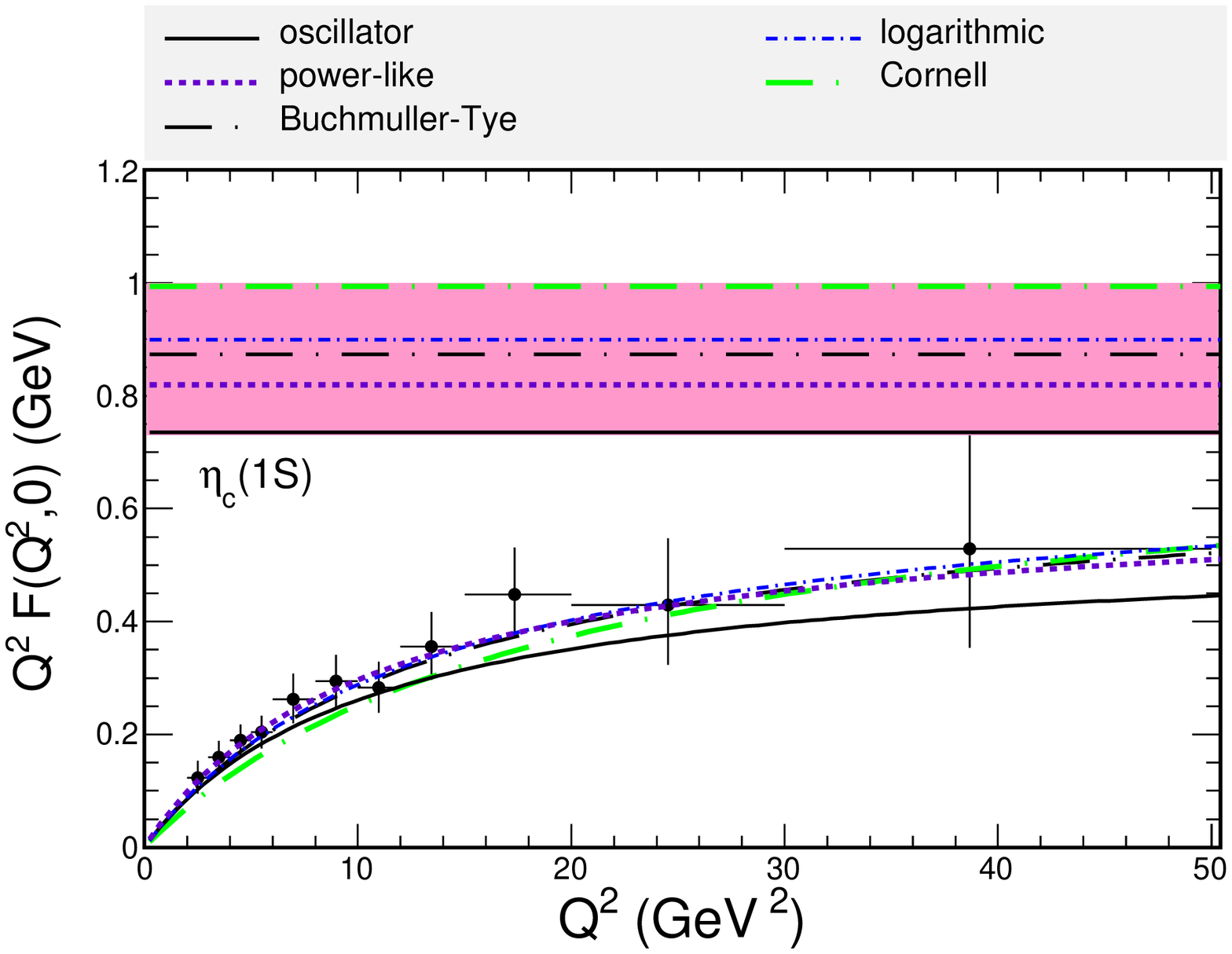}
\includegraphics[width=.7\textwidth]{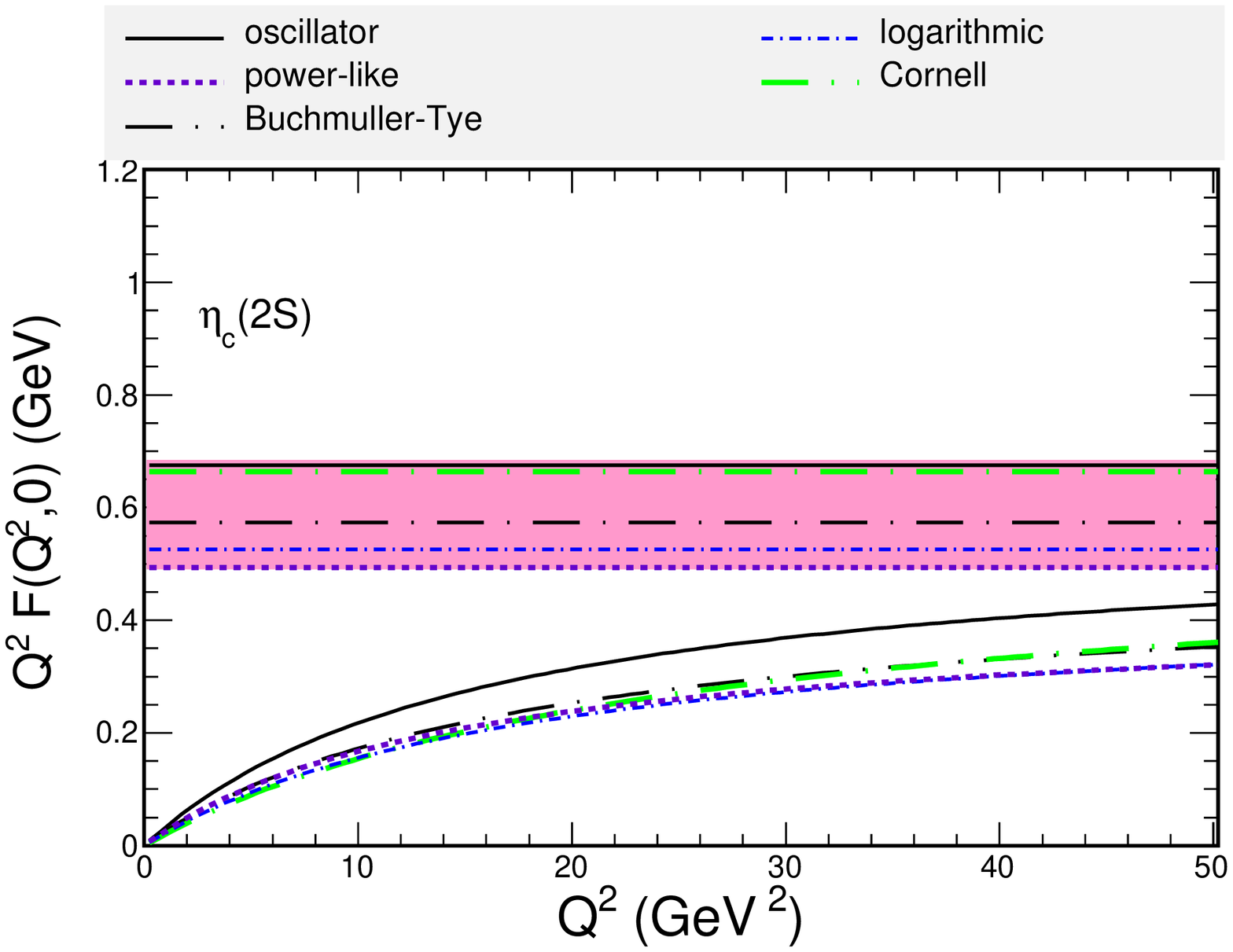}\\
\caption{$Q^2 F(Q^2)$ as a function of photon virtuality for
  $\eta_c$(1S) (upper panel), $\eta_c(2S)$ (lower panel) for different
potentials used in the present paper.}

\label{fig:Q2_FF_Q2}
\end{figure}

In Fig.\ref{fig:Q2_FF_Q2} we show the rate of approaching
of $Q^2 F(Q^2)$ to its asymptotic value predicted by Brodsky and Lepage \cite{Lepage:1980fj}.
For the asymptotic distribution amplitude $\varphi(z,\mu^2) = 6 z(1-z)$, one
would obtain $Q^2 F(Q^2) = \frac{8}{3} f_{\eta_c}$.
Therefore the horizontal lines $\frac{8}{3}f_{\eta_{c}}$ are shown for reference
(upper panel - $\eta_c$(1S), lower panel - $\eta_c$(2S)), considering the values for $f_{\eta_{c}}$ presented in Tables \ref{tab:transition_form_factor_1S} and \ref{tab:transition_form_factor_2S}.
We do not observe approaching towards BL asymptotic value for $Q^2 \le 50$ GeV$^2$. While our results flatten out at large $Q^2$, their asymptotic value is
much smaller then the one predicted within the hard scattering formalism.

%
\begin{figure}
    \centering
    \includegraphics[width = .7\textwidth]{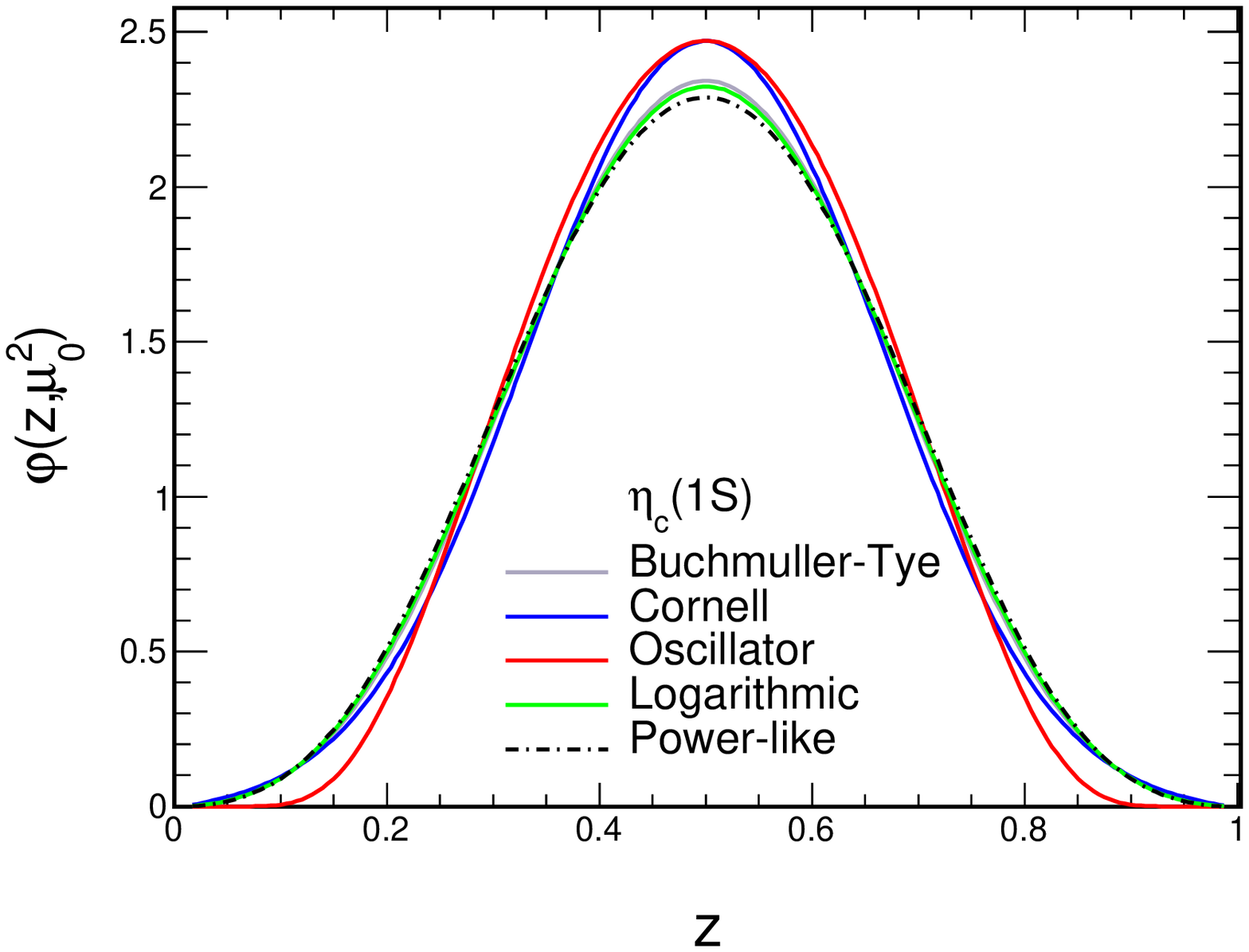}
    \includegraphics[width = .7\textwidth]{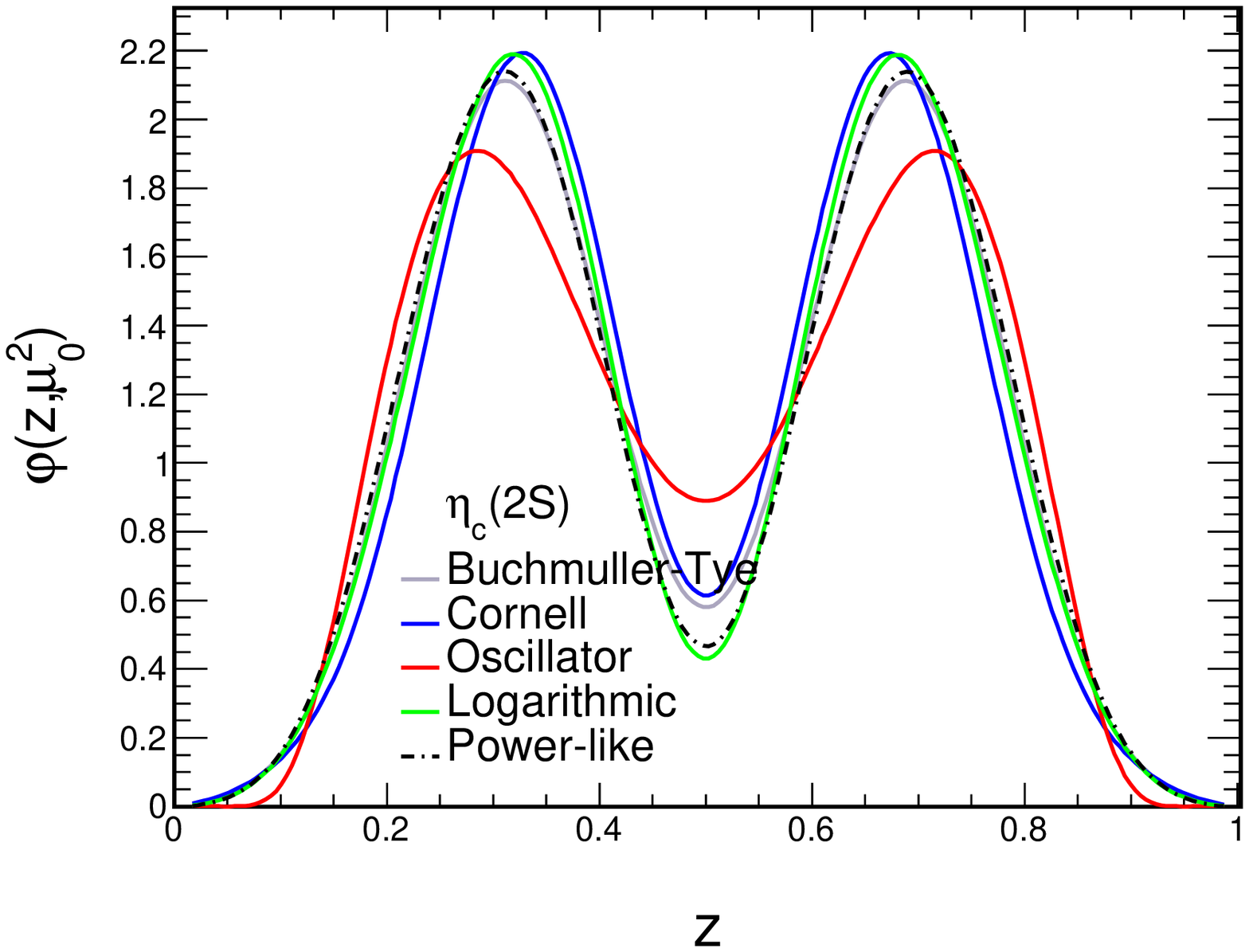}
    \caption{Distribution amplitudes for different wave functions for
      $\eta_c$ (1S) (upper panel) and for $\eta_c$ (2S) (lower panel).}
    \label{fig:DA}
\end{figure}

In order to understand the results for $Q^2 F(Q^2)$ shown above, let us discuss the  applicability of collinear  approach,  commonly used for light pseudoscalar mesons,  for the $\eta_c$ case. 
From Eq.~(\ref{eq:collinear_FF}), one has that the collinear formula with the correct  
asymptotics reads:
\begin{equation}
F(Q^2,0) = F(0,Q^2) =
e_c^2 f_{\eta_c}  \int_0^1 dz\,  \Big\{ {z\,\varphi(z,\mu^2)
\over z^2 Q^2 + m_c^2}  
+ {(1-z)\varphi(z,\mu^2) \over (1-z)^2 Q^2 + m_c^2} 
\Big\} \; .
\label{collinear_formula}
\end{equation}
The evolution with the factorization scale 
is easily implemented in the formalism of distribution amplitudes.
This is routinely  done for light pseudoscalar mesons ($\pi^0 \, , \eta\,
, \eta'$) see e.g. a recent NLO analysis \cite{KPK2019}.
For $c\bar{c}$ (or $b \bar{b}$) quarkonia the situation is much more
complicated and quark mass effects and/or higher twists must be included.
In Fig.~\ref{fig:DA} we show the distribution amplitude at a factorization
scale $\mu = 3 \, \rm{GeV}$ calculated from our
wave functions for both $\eta_c$ (1S) and $\eta_c$ (2S).
To perform the evolution with the hard scale, the distribution amplitude is 
expanded with the help of the Gegenbauer $C^{3/2}_n$ polynomials:
\begin{equation}
\varphi(z,\mu^2) = 6 z (1-z) \left( 1 + a_2(\mu^2) C_2^{3/2}(2z-1) + ... \right) \; .
\label{distribution_amplitude}
\end{equation}
We extract the Gegenbauer coefficients 
by means of
\begin{eqnarray}
a_n(\mu_0) = \frac{2(2n+3)}{3(n+1)(n+2)}\cdot \int_0^1{dz \varphi(z,\mu_0)  C_n^{3/2}(2z-1)} \, .
\end{eqnarray}
They evolve according to
\begin{eqnarray}
a_n(\mu)& = a_n(\mu_0)\cdot \left[ \frac{\alpha_s(\mu)}{\alpha_s(\mu_0)} \right]^{\gamma_n/\beta_0} \, ,
\label{eq:a_n}
\end{eqnarray}
with the the anomalous dimensions $\gamma_n$, which can be found for example in Ref. \cite{Lepage:1980fj}.

For $\eta_c(1S)$, the $a_2$ coefficient dominates and is typically -0.3 at the initial evolution scale.
For the $\eta_c(2S)$ the $n=4,6$ coefficients dominate. We show the Gegenbauer coefficients at the scale $\mu_0 = 3 \, \rm GeV$ for both $\eta_c(1S)$ and $\eta_c(2S)$ in Table \ref{table:coeff}.
\begin{table}
\caption{Extracted coefficients $a_n(\mu_0)$, for the Buchm\"uller-Tye potential}
\label{table:coeff}
\begin{tabular}{l  c c}
\hline
\hline
n & $a_n(\mu_0)\;\eta_c(1S)$ & $a_n(\mu_0)\; \eta_c(2S)$\\
\hline
2 &-0.284     &-0.0765\\
4 & 0.0635   &-0.1627\\
6 &-0.008157 &0.128\\
8 &-0.000619 &-0.049\\
10& 0.000216 &0.0088\\
\hline
\hline 
\end{tabular}
\end{table}

\begin{figure}
\includegraphics[width=.6\textwidth]{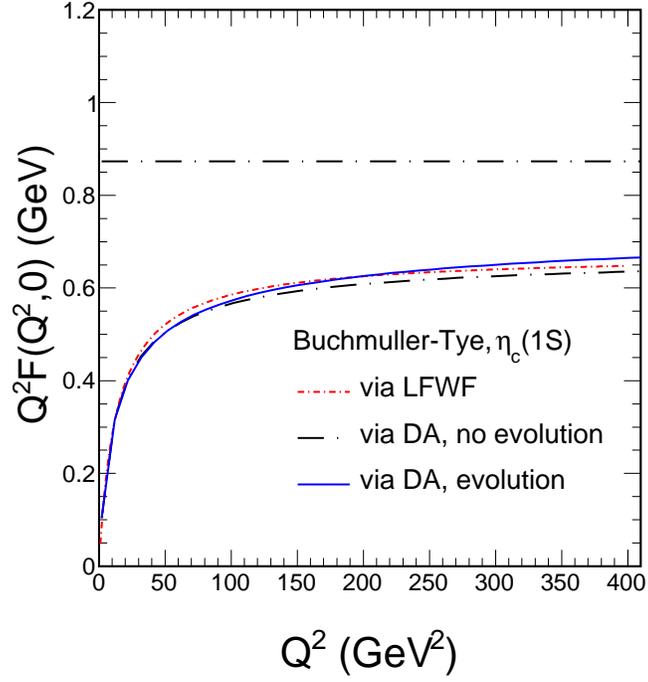}
\includegraphics[width=.6\textwidth]{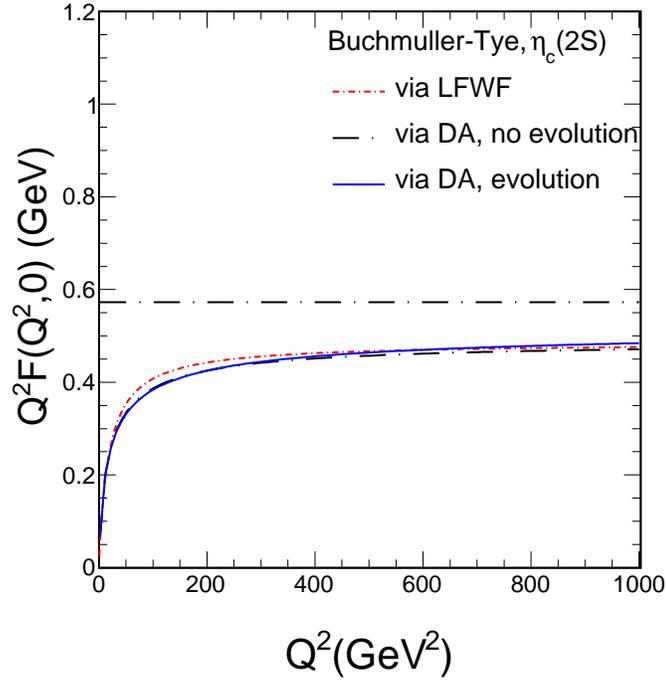}
\caption{
$Q^2 F(Q^2)$ for $\eta_c$ (1S)(upper panel) and  $\eta_c$ (2S) (lower panel) as a function of photon virtuality. The
horizontal line is the limit for $Q^{2}\rightarrow \infty$, calculated 
for the Buchm\"uller-Tye potential. 
The red dashed line (via LFWF) was obtained according to Eq.~(\ref{eq:FF}). The black dashed-dotted curve (via DA) and the blue solid curve (via DA, evolution) were calculated with Eq.~(\ref{collinear_formula}) respectively with and without evolution of Gegenbauer coefficients $a_n$. 
}
\label{fig:collinear}
\end{figure}

In Fig.~\ref{fig:collinear} we illustrate the effect of evolution
on $Q^2 F(Q^2)$. We compare the result obtained with our original 
formula, given by Eq.~(\ref{eq:FF}). In addition, we show results obtained
with the collinear formula (\ref{collinear_formula}) using distribution amplitudes
obtained from our light-front wave functions (see Fig.\ref{fig:DA}). 
There is only some difference at low $Q^2$. 
Finally we show also result obtained within collinear approach 
with QCD evolution of distribution
amplitudes built in, starting from $\mu_0 = 3 \, \rm GeV$.  The effect of evolution is very weak. 
The reader is asked to notice much broader range of $Q^2$ in the
figure compared to that in previous figures. 
Summarizing the effect of evolution can be safely neglected for 
$Q^2 <$ 100 $GeV^2$, i.e. in the range of our interest, i.e. 
where $F(Q^2)$ can be measured.
 
\begin{figure}
\includegraphics[width=6cm]{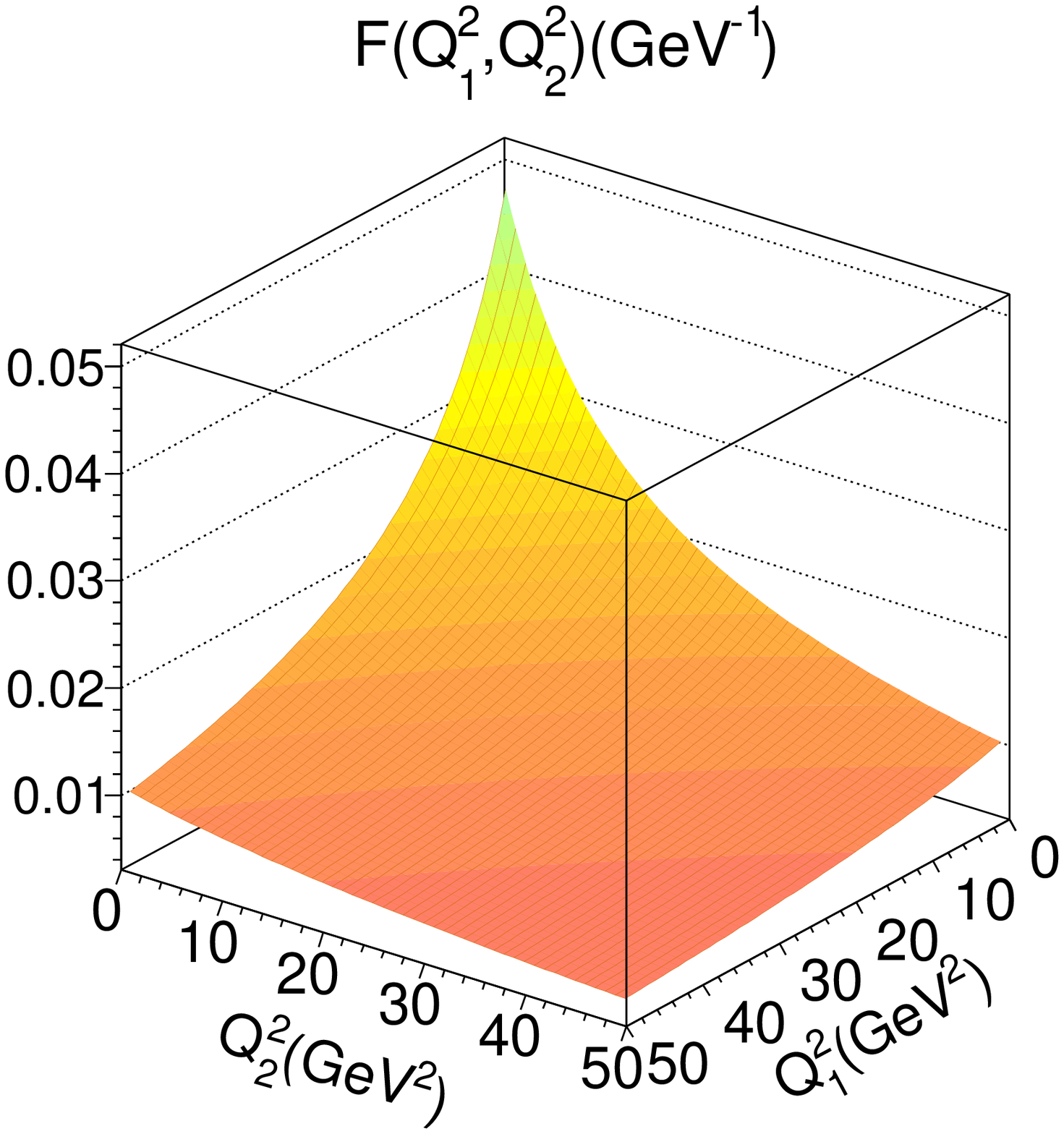}
\includegraphics[width=6cm]{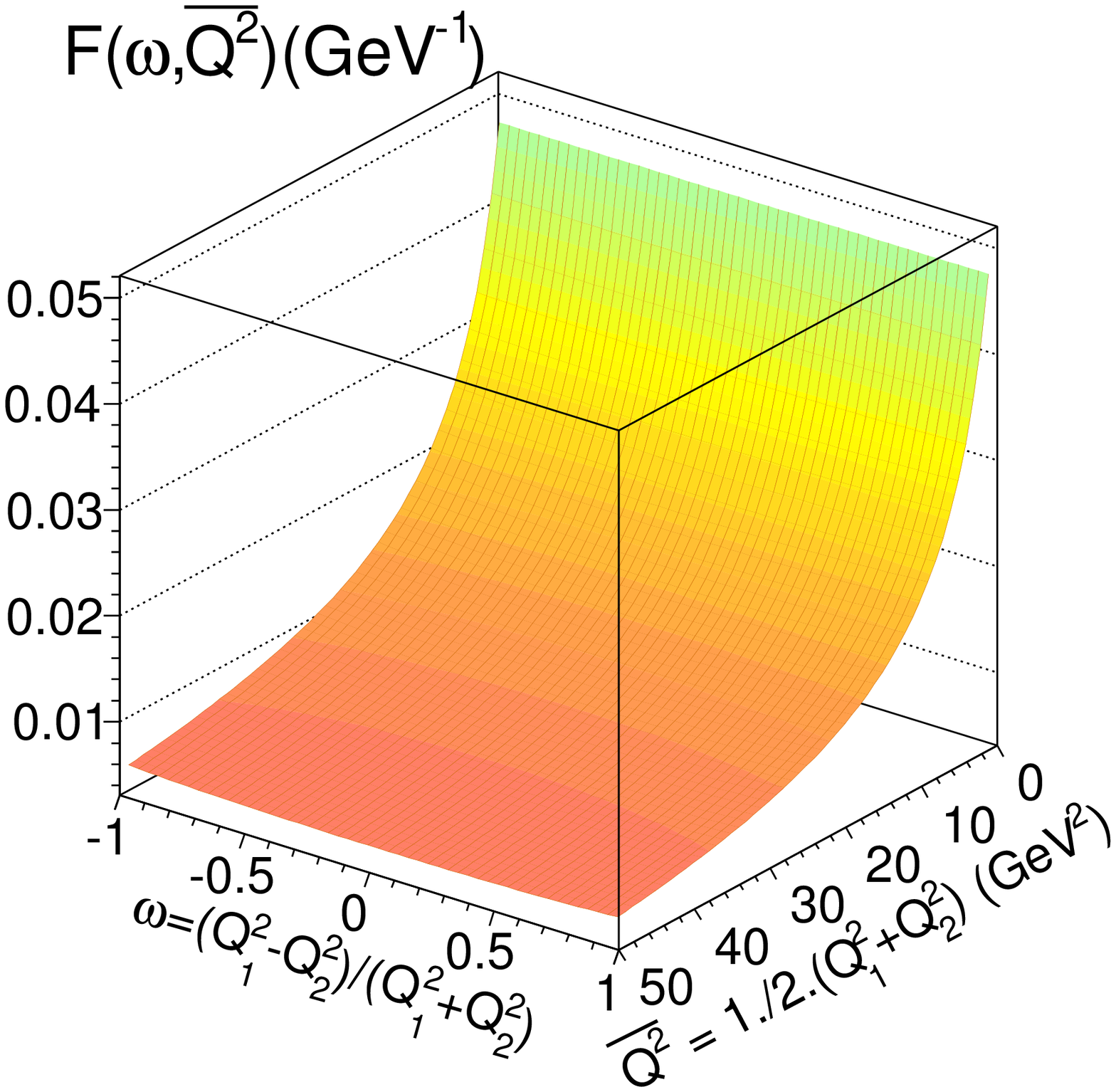}\\
\includegraphics[width=6cm]{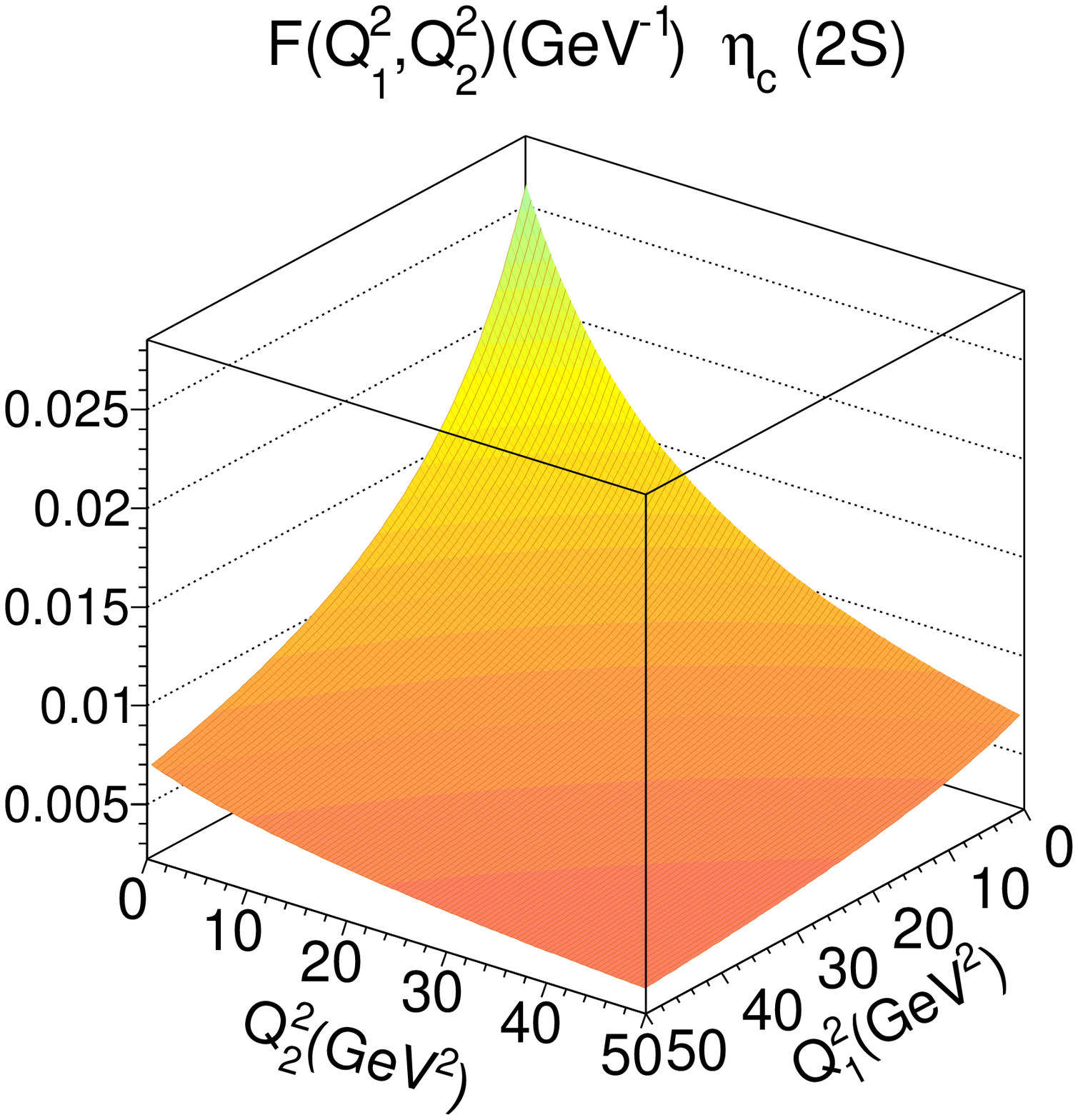}
\includegraphics[width=6cm]{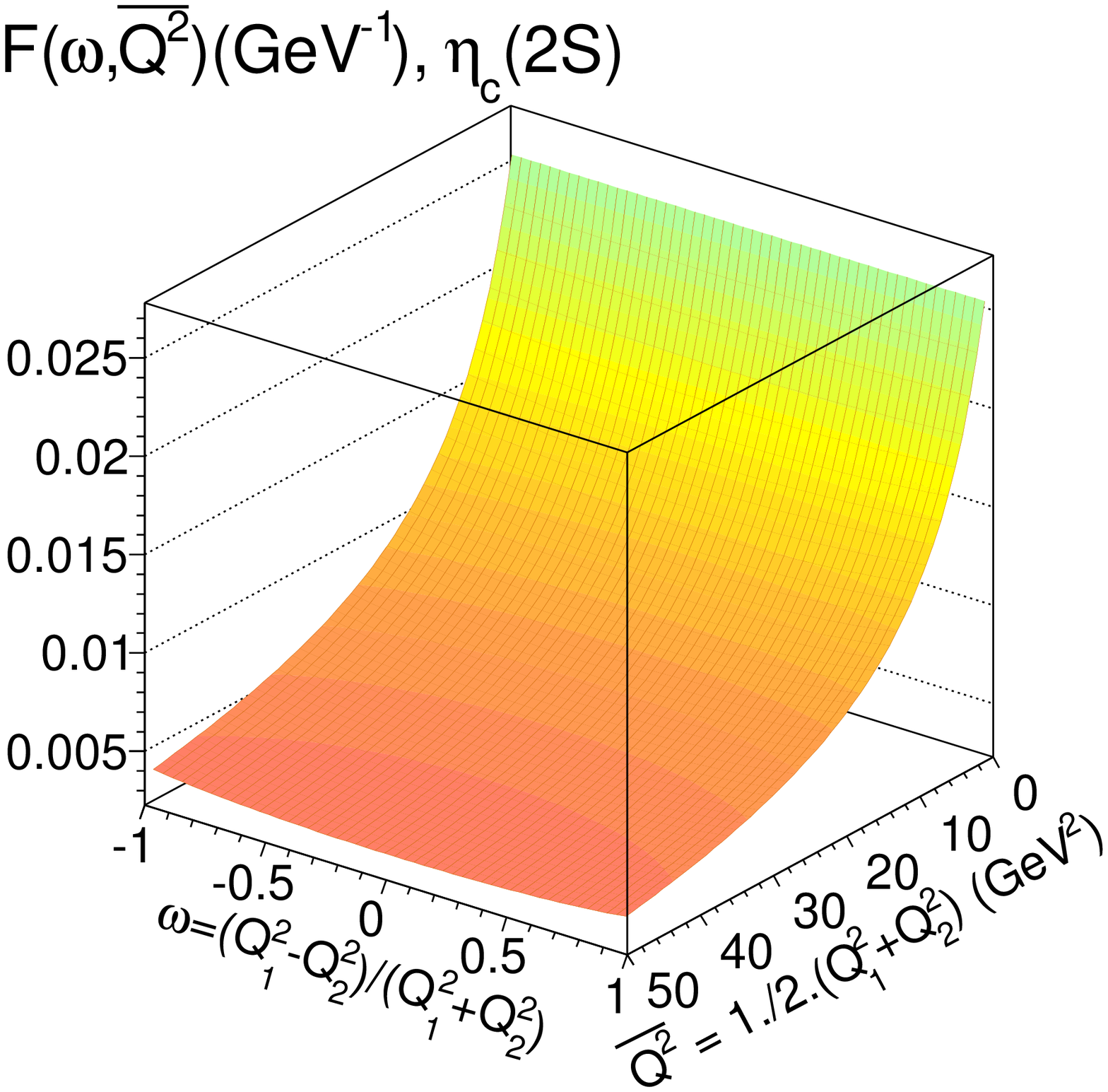}\\

\caption{The $\gamma^*\gamma^*\rightarrow \eta_{c}$ (1S) (top panels) and
             $\gamma^*\gamma^*\rightarrow \eta_{c}$ (2S) (bottom panels) 
form factor as a function of ($Q_1^2,Q_2^2$) and ($\omega,\bar{Q^2}$)
for the Buchm\"uller-Tye potential for illustration, other potentials discussed in the present paper behave similarly.
}
\label{fig:F_Q12Q22}
\end{figure}

Now we wish to present also two-dimensional distributions for the
$\gamma^* \gamma^*$ transition form factor
as a function of the photon virtualities $Q_1^2$ and $Q_2^2$.
As an example in Fig.\ref{fig:F_Q12Q22} we again show our results
for the Buchm\"uller-Tye potential.
To investigate the scaling properties, we show the transition
form factor as a function of the variables 
\begin{eqnarray}
\omega=\frac{Q_1^2-Q_2^2}{Q_1^2+Q_2^2}\, \,\,\,\mbox{and}\,\,\,\bar{Q}^2 = \frac{Q_1^2+Q_2^2}{2}\, .
\end{eqnarray}
One can see that $F$ (and $\widetilde{F}$) is almost  independent 
of the asymmetry parameter $\omega$.
For comparison $\gamma^*\gamma^*\to \pi^0$ transition the dependence 
on $\omega$ is somewhat stronger \cite{DKV2001}.
Note that for the VDM model (\ref{FF_VDM}) some dependence on $\omega$
would be obtained. 
Future investigation of the slow $\omega$ dependence
would be in our opinion an interesting task for Belle 2.

\begin{figure}
\includegraphics[width=6cm]{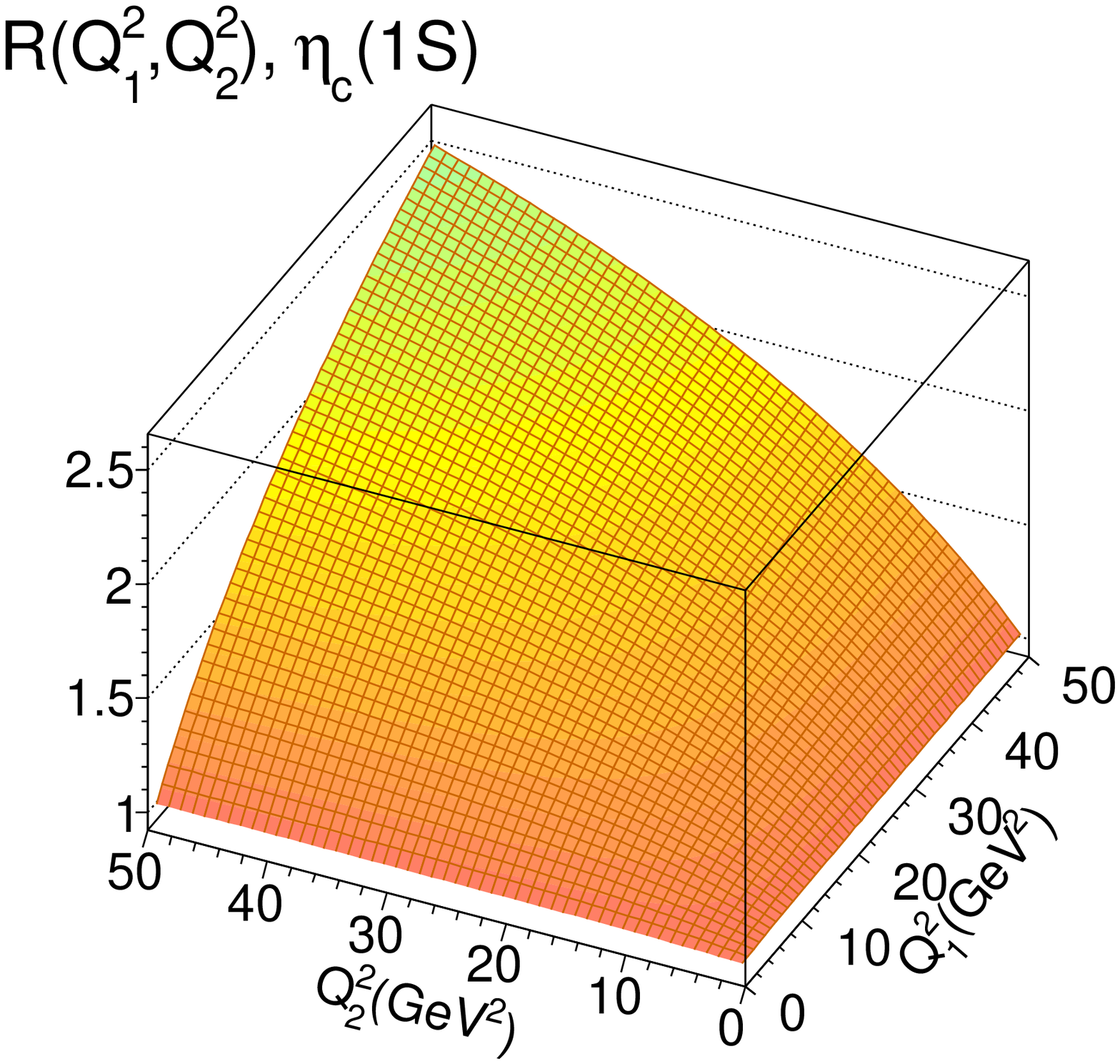}
\includegraphics[width=6cm]{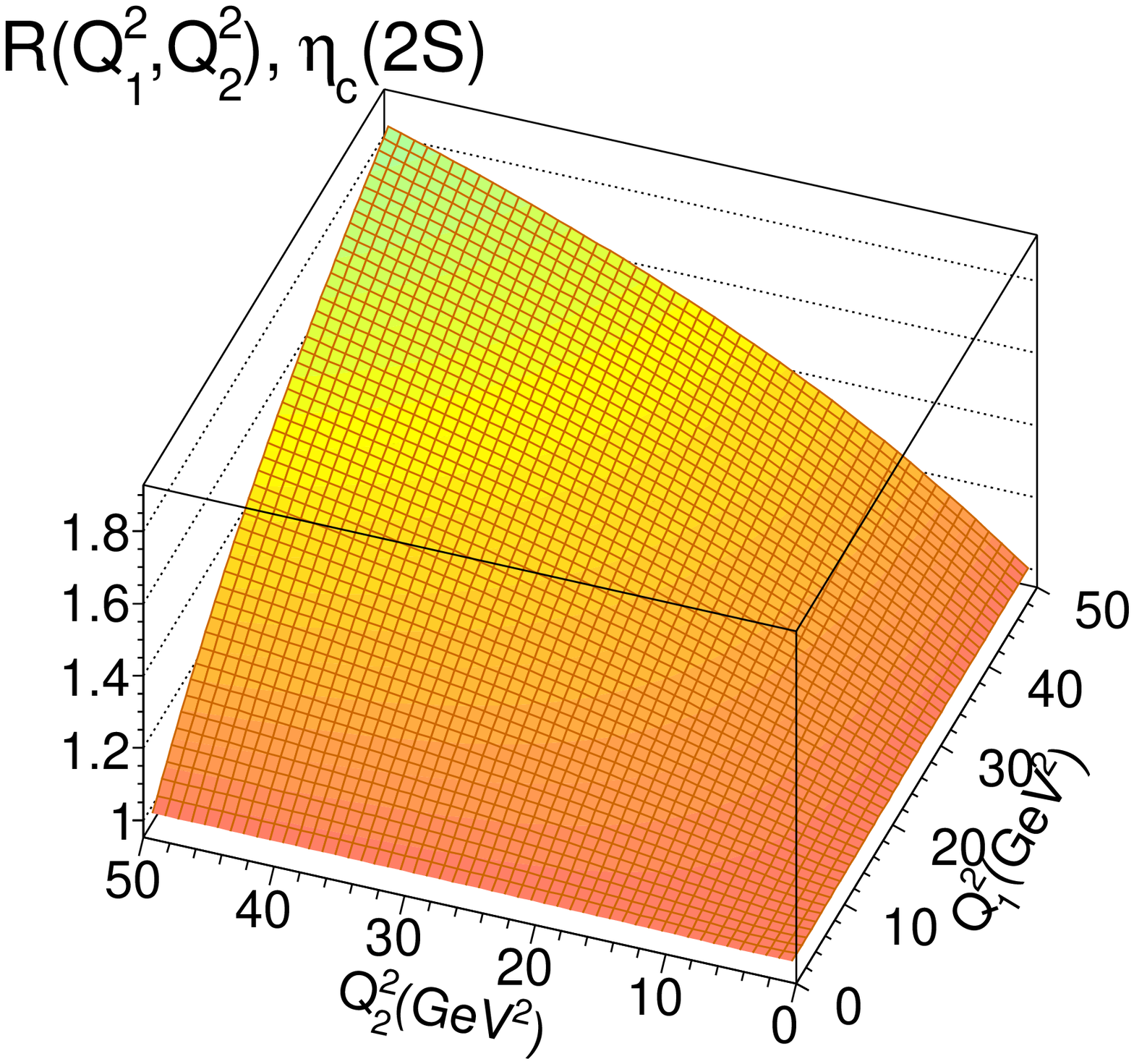}\\
\caption{The deviations from the factorization breaking ($R$) as 
a function of ($Q_1^2,Q_2^2$) for Buchm\"uller-Tye potential; left panel
- $\eta_{c}$(1S), right panel - $\eta_{c}$(2S)).}
\label{fig:R_Q12Q22}
\end{figure}

In Fig.~\ref{fig:R_Q12Q22} we show deviations from the factorization
breaking (see Eq.~(\ref{factorization_breaking})), for 
the Buchm\"uller-Tye potential.
We observe that $R(0,Q_2^2) = R(Q_1^2,0)= 1$. The factorization breaking
pattern looks very similar for different potentials (not shown
explicitly here).

\section{Conclusions}
\label{Sect:Conclusions}

The description of transition form factors is directly related to our understanding of the structure of bound states in QCD. In the present paper we have studied the transition form factors
for $\gamma^* \gamma^* \to \eta_c$ (1S,2S) for two space-like virtual photons, which can be accessed experimentally in future measurements of the cross section for the $e^+ e^- \rightarrow e^+ e^- \eta_c$ process in the double - tag mode. The light-front wave function representation of these observables
has been derived and discussed.

We have calculated the transition form factor for different
wave functions obtained as a solution of the Schr\"odinger equation for the $c \bar c$ system for different phenomenological
$c \bar{c}$ potentials from the literature. 
The rest-frame momentum space wave functions
have been transformed to the light-front representation
using the Terentev prescription. 
Firstly we have presented the transition form factor for only one
off-shell photon as a function of its virtuality and compared
to the BaBar data for the $\eta_c(1S)$ case.
We have also presented the delayed convergence of the form factor to
its asymptotic value $\frac{8}{3}f_{\eta_{c}}$ as predicted by 
the standard hard scattering formalism.
Our results for $Q^2F(Q^2)$ approach to a lower asymptotic value.
The Brodsky-Lepage limit can only be obtained after 
including QCD evolution of the distribution amplitudes 
for massless quarks, but appears irrelevant in the accessible kinematic domain. 
We conclude that it is not necessary to include the QCD evolution 
for $Q^2 <$ 100 GeV$^2$. This justifies, a posteriori, our results
obtained within the approach using $c \bar c$ wave functions.
Furthermore, we have presented two-dimensional distributions 
in the virtualities of photons
of the $\gamma^* \gamma^* \eta_c$ transition form factor for $\eta_c(1S)$ and $\eta_c(2S)$.
We have predicted a very slow dependence on the asymmetry parameter 
$\omega$, which 
could be verified experimentally at Belle 2.
We have also defined a measure of factorization breaking and have calculated
it for different potentials as a function of $(Q_1^2,Q_2^2)$. 
The results on the $Q_1^2, Q_2^2$ dependence are almost model independent.

Finally, it is important to emphasize that the $\eta_c$ production in hadronic collisions is dominated, at lowest order, by the $g^* g^* \rightarrow \eta_c$ subprocess, which is identical (up to global color factors) to the
$\gamma^* \gamma^* \rightarrow \eta_c$ amplitude derived in this paper. Consequently, our results can also be useful to estimate the $\eta_c$ production at the LHC, which is currently a theme of intense debate (see e.g. Refs. \cite{kniel_prl15,han_prl15,zhang_prl15}).

\section{Acknowledgments}

V.P.G. was  partially financed by the Brazilian funding agencies CNPq, 
FAPERGS and INCT-FNA (project number 464898/2014-5). R.P.~is supported 
in part by the Swedish Research Council grants, contract numbers 621-2013-4287 
and 2016-05996, by CONICYT grant MEC80170112, as well as by the European Research Council (ERC) under the European Union's Horizon 2020 research and innovation 
programme (grant agreement No 668679). This work was also supported in part by the Ministry of Education, Youth and Sports of the Czech Republic, project LT17018. The work has been performed in the framework of COST Action CA15213 
``Theory of hot matter and relativistic heavy-ion collisions'' (THOR). 


\end{document}